\documentclass[superscriptaddress,reprint,amsmath,amssymb,aps,prb]{revtex4-1}
\usepackage{graphicx}
\usepackage{dcolumn}
\usepackage{bm}
\usepackage{color}
\usepackage{ulem}

\begin{document}
\title{Magnetic clustering, half-moons, and shadow pinch points as 
signals of a\\ proximate Coulomb phase in frustrated Heisenberg magnets}
\author{Tomonari Mizoguchi}
\affiliation{Department Physics, Gakushuin University, Mejiro, Toshima-ku, Tokyo 171-8588, Japan}
\email{mizoguchi@rhodia.ph.tsukuba.ac.jp}
\altaffiliation[Present address: ]{Department of Physics, University of Tsukuba,
Tsukuba, Ibaraki 305-8571, Japan}
\author{Ludovic D. C. Jaubert}
\affiliation{Universit\'e Bordeaux, CNRS, LOMA, UMR 5798, F-33405 Talence, France}
\affiliation{Okinawa Institute of Science and Technology Graduate University, Onna-son, Okinawa 904-0495, Japan}
\author{Roderich Moessner}
\affiliation{Max-Planck-Institut fur Physik komplexer Systeme, Nothnitzer Str. 38, 01187 Dresden, Germany}
\author{Masafumi Udagawa}
\affiliation{Department Physics, Gakushuin University, Mejiro, Toshima-ku, Tokyo 171-8588, Japan}
\date{\today}
\begin{abstract}
We study the formation of magnetic clusters in frustrated magnets in their cooperative paramagnetic regime. For this purpose, we consider the $J_1$-$J_2$-$J_3$ classical Heisenberg model on kagome and pyrochlore lattices with $J_2 = J_3=J$. 
In the absence of farther-neighbor couplings, $J=0$, the system is in the Coulomb phase with magnetic correlations well characterized by pinch-point singularities. Farther-neighbor couplings lead to the formation of magnetic clusters, which can be interpreted as a counterpart of topological-charge clusters in Ising frustrated magnets [T. Mizoguchi, L. D. C. Jaubert and M. Udagawa, Phys. Rev. Lett. {\bf 119}, 077207 (2017)]. 
The concomitant static and dynamical magnetic structure factors, respectively $\mathcal{S}({\bm{q}})$ and $\mathcal{S}({\bm{q}},\omega)$, develop half-moon patterns. As $J$ increases, the continuous nature of the Heisenberg spins enables the half-moons to coalesce into connected ``star'' structures spreading across multiple Brillouin zones. These characteristic patterns are a dispersive complement of the pinch point singularities, and signal the proximity to a Coulomb phase. Shadows of the pinch points remain visible at finite energy, $\omega$. This opens the way to observe these clusters through (in)elastic neutron scattering experiments.
The origin of these features are clarified by complementary methods: large-$N$ calculations, semi-classical dynamics of the Landau-Lifshitz equation, and Monte Carlo simulations. 
As promising candidates to observe the clustering states, we revisit the origin of ``spin molecules" observed in a family of spinel oxides  $AB_2$O$_4$ ($A=$ Zn, Hg, Mg, $B=$ Cr, Fe). 

\begin{description}
\item[PACS numbers] 75.10.Kt
\end{description}
\end{abstract}

\pacs{Valid PACS appear here}

\maketitle
\section{Introduction}
Geometrically frustrated magnets provide a stage to realize exotic states of matter, ranging from quantum and classical spin liquids~\cite{balents2010, savary2016, zhou2017}, unconventional magnetic ordering with topological response~\cite{taguchi2001,onose2010,nagaosa2013}, and states accompanied by exotic phase transitions~\cite{,kawamura_miyashita1984,weber2003, jaubert2008}.
Among them, the disordered Coulomb phase is a canonical example for discrete and continuous spins on the three-dimensional pyrochlore lattice  when frustration imposes a local divergence-free constraint~\cite{henley2010}, whose exotic character has been drawing considerable interest.

The Coulomb phase is based on degenerate classical spin configurations in absence of any spontaneous symmetry breaking. The magnetic correlations due to the local divergence-free constraint are characterized by non-analyticities in the static magnetic structure factor, $\mathcal{S}({\bm{q}})$~\cite{isakov2004}. These are called pinch points, and have been observed in the canonical spin-ice materials, Ho$_2$Ti$_2$O$_7$ and Dy$_2$Ti$_2$O$_7$~\cite{bramwell2001,fennell2009}.

On top of exotic correlations, the Coulomb phase supports fractional excitations.
These excitations are easy to visualize in spin ice, where Ising spins satisfy the so-called ice rules with two spins pointing inwards and two spins pointing outwards on every tetrahedron in the ground state.
A tetrahedron in a ``three-in-one-out" or ``one-in-three-out" configuration, breaking this ice rule, carries a gauge charge and serves as an elementary fractional excitation. In spin ice, these gauge charges are actually effective magnetic charges~\cite{castelnovo2008}. By identifying the spins with their inherent ``magnetic field," one can regard the gauge charge as a source or sink of the field, and assign a magnetic charge $-2 (+2)$ 
for  ``one-in-three-out" (``three-in-one-out") tetrahedron states. Magnetic charges are defined from the discrete divergence of the magnetic field, i.e., the number of inward spins minus that of outward spins.

The introduction of the concept of magnetic charges turned out to be quite illuminating, carrying over two properties from conventional electromagnetism. Firstly, charge should be conserved. Indeed, the above-mentioned magnetic charge in spin ice satisfies a local conservation in the sense that they are always created/annihilated in pairs of positive and negative charges. And, secondly, opposite charges are expected to attract each other. However, this second property is non-universal. In the canonical spin-ice systems, Ho$_2$Ti$_2$O$_7$ and Dy$_2$Ti$_2$O$_7$, opposite charges indeed interact with attractive force, attributed to the long-range dipolar interaction. However, the sign of the force actually depends on the microscopic details of the system.

Indeed, recently, the role of charge interactions is drawing interest in spin ice~\cite{udagawa2016,rau2016}
and its two-dimensional analog~\cite{moller2009,chern2011,chern2012,mizoguchi2017}. If the interaction is chosen ``unnaturally," i.e., attractive between same-sign charges, the Coulomb phase is destabilized towards the formation of same-sign-charge hexamer clustering~\cite{udagawa2016,rau2016,mizoguchi2017}. The generic tendency to clustering can be naturally understood from the competition of the two-fold properties of charges. Same-sign charges attract each other, but they cannot pair-annihilate due to charge conservation. As a result, they form stable clusters. Their proliferation leads to unconventional classical spin liquids. Accompanying the clustering, the magnetic correlations display a noticeable evolution, characterized by half-moon patterns in $\mathcal{S}({\bm{q}})$, which replace the pinch point singularities~\cite{robert2008,udagawa2016,rau2016,mizoguchi2017}.

The rich physics brought by the attraction of charges of the same sign naturally motivates us to generalize its analysis to the system with continuous spins. Indeed, the magnetic charge in the Ising system can be generalized to a conserved magnetic vector in continuous spin systems. With the continuous nature of magnetic vector, one can expect a variety of stable textures beyond hexamer clustering.
From this viewpoint, it is interesting to look at a class of spinel oxides with 3$d$ magnetic ions, $AB_2$O$_4$ ($A=$ Zn, Hg, Mg, $B=$ Cr, Fe)~\cite{lee2002,tomiyasu2011,tomiyasu2008,tomiyasu2013,gao2018,yamada2002,kamazawa2003,conlon2010,tomiyasu2011_2}.
These compounds have weak magnetic anisotropy with small spin-orbit interaction of $3d$ ions, and the classical Heisenberg model with farther-neighbor interactions is expected to give a good starting point of analysis~\cite{conlon2010}.
Indeed, according to inelastic neutron scattering experiments, this family lacks ``pinch points" in the dynamical structure factor, which are characteristic of the Coulomb phase. Instead, diffuse scattering patterns appear at the corners of the Brillouin zone.
The diffuse scatterings are attributed to the clustering of small number of spins, coined as ``spin molecules."
Depending on materials, molecules take the form of hexamers~\cite{lee2002,tomiyasu2011,tomiyasu2008,tomiyasu2013,gao2018} and dodecamers~\cite{yamada2002,kamazawa2003,tomiyasu2011_2}.
In the work by Conlon and Chalker \cite{conlon2010}, the lack of pinch points has been attributed to weak, generic, farther-neighbor exchange, inducing hexagonal cluster scattering as observed in experiments.

To address these issues, in this paper, we consider the classical Heisenberg models on kagome and pyrochlore lattices with farther-neighbor interactions, on the high-symmetry line $J_2=J_3=J$, for arbitrary values of $J>0$. We focus on cooperative paramagnetic region above magnetic ordering temperature, where the magnetic fluctuations reflect the intrinsic nature of the system, in contrast to the ordering pattern itself, which is susceptible to structural changes or other extrinsic effects.

Our main results are summarized as follows:
(i) We found three distinct patterns in $\mathcal{S}(\bm{q})$: pinch points, half-moons, and stars.
These patterns are counterparts of the topological-charge clusters obtained in the corresponding Ising models. 
(ii) The three patterns reflect the structure of softest magnetic modes.
(iii) The half-moon and star patterns can be interpreted as shadows of pinch points, and serve as a signal of proximity to a Coulomb phase.
(iv) These characteristic patterns also appear in the low-energy region of dynamical structure factors, implying the possibility of experimental detection through inelastic neutron scattering.

\begin{figure}[b]
\begin{center}
\includegraphics[bb = 0 0 496 294, width=7.5cm]{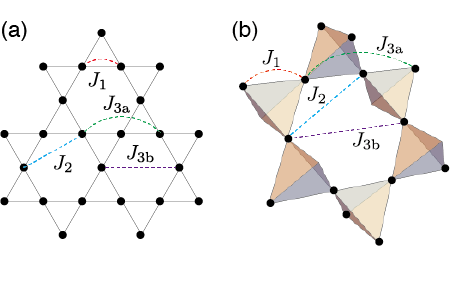}\\
\vspace{-10pt}
\caption{Schematic picture of the model in Eq. (\ref{eq:hami}) for (a) a kagome lattice and (b) a pyrochlore lattice. 
Red,  blue, green and purple lines denote, respectively, $J_1$, $J_2$, $J_{3a}$ and $J_{3b}$. 
In this paper, we consider $J_{1}=1$, $J_{2}=J_{3a}=J$ and $J_{3b}=0$.}
\label{fig:bandkagome}
\end{center}
\end{figure}

The rest of this paper is organized as follows.
In Sec.~\ref{sec:model_and_formalism}, we first describe the model, namely, the $J_1$-$J_2$-$J_3$ classical Heisenberg Hamiltonian on the kagome and pyrochlore lattices. 
Next, we introduce the theoretical methods; 
large-$N$ calculations, Monte Carlo simulations, and semiclassical Landau-Lifshitz (LL) equation.
In Sec.~\ref{sec:ssf}, the three distinct patterns of the static structure factor and their origins are discussed. 
Here, the main arguments are based on band-structure analyses of the large-$N$ approximation, supported by Monte Carlo simulations on the O(3) Heisenberg model. 
Section~\ref{sec:semiclassical_dynamics} is devoted to the spin dynamics, analyzed by the LL equation. 
In Section~\ref{sec:ising}, we discuss the real-space structure of the clusters, and show that they can be understood as a continuation from the topological charge cluster obtained in the Ising counterpart.
Finally, we present discussions and summary in Sec.~\ref{sec:summary}. 
Details of the large-$N$ approximation, Monte Carlo simulations,
and the quadrupolar order parameter
are described in the Appendices. 

\section{Model and formalism}
\label{sec:model_and_formalism}

\subsection{Model}
We consider a Heisenberg model on kagome and pyrochlore lattices with up to third-neighbor interactions:
\begin{align}
H = & J_1 \sum_{\langle i,j\rangle_{\mathrm{NN}} } \bm{S}_i \cdot \bm{S}_j + J_2 \sum_{\langle i,j\rangle_{\mathrm{2nd}}} \bm{S}_i \cdot \bm{S}_j  \notag \\
+ & J_{3\mathrm{a}} \sum_{\langle i,j\rangle_{\mathrm{3rd,a}}} 
\bm{S}_i \cdot \bm{S}_j +J_{3\mathrm{b}} 
\sum_{\langle i,j\rangle_{\mathrm{3rd,b}}} \bm{S}_i \cdot \bm{S}_j \notag \\
= & \frac{1}{2}\sum_{n,m}\sum_{\mu,\nu} \sum_{\alpha = x,y,z}
S_n^{\mu,\alpha} H_{n,m}^{\mu,\nu} S_m^{\nu,\alpha}.
\label{eq:hami}
\end{align}

$J_1$, $J_2$, and $J_{3\mathrm{a}, 3\mathrm{b}}$ are the exchange coupling constants connecting nearest-, second-nearest-, and third-nearest neighbors, as shown in Fig.~\ref{fig:bandkagome}.
Notice that two different types of the third-neighbor term ($J_{3\mathrm{a}}$ and $J_{3\mathrm{b}}$)
are distinguished in that $J_{3\mathrm{a}}$ connects the sites along edges, 
while $J_{3\mathrm{b}}$ connects those across a hexagon. 
Here, we have introduced unit cell indices, $n$, $m$, and sublattice indices, $\mu$, $\nu$, and expressed each site as their combinations: $i=(n,\mu)$ and $j=(m,\nu)$.
The unit cells contain respectively triangles (tetrahedra) of one orientation for kagome (pyrochlore) lattice, and the sublattice indices take $\mu= 1, \cdots N_{\rm sub}$, with $N_{\rm sub}=3$ (4).
$\bm{S}_n^{\mu} = (S_n^{\mu,x}, S_n^{\mu,y} ,S_n^{\mu,z})$ 
are (classical) three-component vectors with unit length $|\bm{S}_n^\mu|  =1$. 

The model with a general parameter set has been intensively studied on both kagome~\cite{chalker1992,huse1992,reimers1993,zhitomirsky2008,chern2013,albaracin2018,li2018} and 
pyrochlore~\cite{reimers1991,moessner1998,canals2001,chern2008,conlon2010,okubo2011,iqbal2018} lattices, 
putting an emphasis on the low-temperature ordered states. 
In the present work, we focus on the parameters $(J_1, J_2, J_{3\mathrm{a}}, J_{3\mathrm{b}}) = (1,J,J,0)$ with $J>0$. 
The value of $J_{1}=1$ sets the energy and temperature scales of our problem.
This parameter set has recently been shown to lead to a clustering of topological charges in the corresponding Ising models for $J>0$ (see Sec. \ref{sec:ising} for details~\cite{udagawa2016,rau2016,mizoguchi2017}). 

Although $J_2=J_3(=J)$ is unlikely to be perfectly satisfied in real materials, the analysis of this isotropic point gives a great insight into the nearby systems.
As we will show in the next section, this point allows a rewriting of the Hamiltonian with conserved spins, and simplifies the formulation of the large-$N$ analysis (see Sec. \ref{sec:largeNformalism} and Appendix \ref{sec:largeNanalytics}).
These properties make it easier to grasp the physics of half-moons and magnetic clustering, which are useful to understand the nature of realistic systems around this point.

\subsection{Conserved spins}
It is instructive to introduce a local magnetic moment for each triangular and tetrahedral unit, $n$,
\begin{align}
{\mathbf M}_n\equiv\zeta_n\sum_{j \in n } {\bm S}_{j},
\label{eq:Mp}
\end{align}
where $\zeta_n=\pm 1$ is a sign factor distinguishing between upward ($+1$) and downward ($-1$) 
triangles/tetrahedra~\cite{cap1}. 
Under the condition: $J_2=J_{3a}(=J)$ and $J_{3b}=0$, 
the Hamiltonian (\ref{eq:hami}) can be rewritten as a function of ${\mathbf M}_n$:
\begin{align}
H = \left(\frac{1}{2} - J \right) \sum_n|{\mathbf M}_n|^2 - J\sum_{\langle n,m\rangle}{\mathbf M}_n\cdot {\mathbf M}_m,
\label{Ham_conservedspin}
\end{align}
where the summation over $n$ is taken over both upward and downward triangles/tetrahedra, and the summation over $\langle n,m\rangle$ is over neighboring pairs of triangles/tetrahedra.
This expression naturally accounts for the Coulomb phase at $J=0$, with ${\mathbf M}_n=0$ for all $n$, and for its stability for small $J$ as will be discussed in detail in the next sections. Equation (\ref{Ham_conservedspin}) is a generalization of the spin to charge mapping of the corresponding Ising systems~\cite{ishizuka2013, udagawa2016, rau2016, mizoguchi2017} and satisfies a conservation law:
\begin{align}
\sum_{n\in D}{\mathbf M}_n = \sum_{j\in\partial D}{\bm S}_{j}\, ,
\end{align}
where $D$ is a connected ensemble of triangles/tetrahedra, and $\partial D$ is its contour. The contour $\partial D$ is made of all spins shared between two triangles/tetrahedra, $n\in D$ and $m\notin D$.
This ``Gauss' law" means that the internal structure of a magnetic cluster is constrained by its boundary spins. Indeed, in the Ising case, the discrete variant of this Gauss' constraint strictly determines the structure of clusters, and leads to hexamer spin liquids \cite{mizoguchi2017}.

\subsection{Formalism}
We study the static properties of model (\ref{eq:hami}) with $(J_1, J_2, J_{3\mathrm{a}}, J_{3\mathrm{b}}) = (1,J,J,0)$ and $J>0$, by combining classical Monte Carlo simulations and analytical large-$N$ method. We also address the dynamics by simulating the semi-classical LL equation. In this section, we introduce the latter two methods, and all details for the classical Monte Carlo simulations will be given in Appendix~\ref{app:MC}.

\subsubsection{Large-$N$ approximation \label{sec:largeNformalism}}
To investigate static structure factors, we employ a large-$N$ approximation~\cite{conlon2010,garanin1999,isakov2004,sen2013}. 
The length of classical Heisenberg spins satisfies a hard constraint $|\bm{S}_n^\mu| = 1$. 
In the large-$N$ method, Heisenberg spins $\bm{S}_n^\mu $ are replaced by soft-spin variables $s_n^\mu$ whose length is constrained on average: 
\begin{align}
\langle (s_n^{\mu})^2  \rangle   = \frac{1}{3}. 
\label{eq:largeNnorm}
\end{align}
Here disordered phases are assumed with $\langle s_n^{\mu} \rangle   =0$. The above constraint (\ref{eq:largeNnorm}) is enforced by
introducing a Lagrange multiplier $\lambda$ which satisfies
\begin{align}
\frac{1}{N_{\mathrm{site}}}\sum_{\bm{q}} \mathrm{Tr}  [\lambda \hat{I} + \beta \hat{H}(\bm{q}) ]^{-1}   = \frac{1}{3}, \label{eq:constraint}
\end{align}
where the sum runs over all wavevectors $\bm q$ in the Brillouin zone and $N _{\mathrm{site}}$ is the total number of sites. 
$\hat{H}(\bm{q})$ represents the Fourier transformation of the exchange matrix:
\begin{align}
[\hat{H}(\bm{q})]_{\mu\nu} =  \sum_{m} H_{0,m}^{\mu\nu} e^{i \bm{q} \cdot (\bm{R}_m + \bm{r}_\nu -\bm{r}_\mu)}\, , \label{eq:hami_matrix}
\end{align}
where ${\bm R}_{m}$ is the position of unit cell $m$ with respect to the reference $0$, and $\bm{r}_{\mu}$ is the position of the sublattice $\mu$ within a unit cell. The static structure factor $\mathcal{S}({\bm{q}})$ in this formalism is given as 
\begin{align}
\mathcal{S}({\bm{q}}) = &  \sum_{\mu, \nu} \langle s^\mu(-\bm{q}) s^\nu(\bm{q}) \rangle \notag = \sum_{\mu, \nu} \left[ \lambda \hat{I} + \beta \hat{H}(\bm{q}) \right]^{-1}_{\mu\nu} \notag \\
= & \sum_{\eta = 1}^{N_{\mathrm{sub}} } \sum_{\mu, \nu} \frac{ [\bm{\psi}^\ast_\eta (\bm{q})]_\mu [\bm{\psi}_\eta (\bm{q})]_\nu }{ \lambda + \beta \varepsilon_\eta(\bm{q}) },  \label{eq:ssf_diag}
\end{align}
where $\varepsilon_\eta(\bm{q})$ and $\bm{\psi}_\eta (\bm{q})$ are, respectively, eigenvalues and eigenvectors 
of $\hat{H}(\bm{q})$ with a band index $\eta$. 
Their calculation can be carried out by using the premedial lattices of kagome and pyrochlore, which are respectively the honeycomb and diamond lattices. 
The main idea of the analytic calculation is to regard the nearest-neighbor (NN) exchange interaction of the original lattice as being mediated by the sites of the premedial lattice located in-between. 
We describe this method in Appendix A.

\subsubsection{Landau-Lifshitz equation}
To investigate the dynamical properties, we numerically solve the following LL equation~\cite{robert2008,conlon2009,taillefumier2014}:  
\begin{equation}
\frac{\partial \bm{S}_i} {\partial t} = - \bm{S}_i \times \bm{H}_{\mathrm{eff},i} , \label{eq:LL}
\end{equation}
where $\bm{H}_{\mathrm{eff},i}$ is an effective magnetic field at site $i$ given as 
\begin{align}
\bm{H}_{\mathrm{eff},i} = & \frac{\partial H}{\partial \bm{S}_i }  \notag \\
= & J_1 \left(\sum_{j: \langle i,j \rangle \in {\mathrm{NN} }} \bm{S}_j 
\right) + J_2 \left(\sum_{j: \langle i,j \rangle \in {\mathrm{2nd} } } \bm{S}_j \right)  \notag \\
+ &J_{3\mathrm{a}}  \left(\sum_{j: \langle i,j \rangle \in  {\mathrm{3rd,a } }} \bm{S}_j
\right)
 +J_{3\mathrm{b}}  \left(\sum_{j: \langle i,j \rangle\in {\mathrm{3rd,b} }} \bm{S}_j
\right).  \label{eq:eff_field}
\end{align}
In our simulation, 
we first prepare the initial states,
which are well thermalized with temperature $T$,
by using single-spin Metropolis updates. 
Then, we solve Eq. (\ref{eq:LL}) by using fourth order Runge-Kutta method.
We confirmed the accuracy of this method by checking that conserved quantities, such as the total energy, stay invariant during the simulation time.
With this method, we compute $\mathcal{S}({\bm{q}},\omega)$ as 
\begin{align}
\mathcal{S}({\bm{q}},\omega) = & \frac{1}{N_t}\sum_{l=0}^{N_{t} } \sum_{n} \sum_{\mu,\nu} 
\langle \bm{S}_0^\mu (0) \bm{S}_{n}^\nu (l\delta t) \rangle_{\rm init} \notag \\
\times & e^{i [\omega l \delta t -\bm{q}\cdot(\bm{R}_n+\bm{r}_\nu -\bm{r}_\mu)] },
\end{align}
where $\langle \cdots \rangle_{\rm init}$ represents the sample average of independently-prepared initial states. Numerical details are given in Appendix \ref{app:MC}.

\section{Fourier-space analysis:\\ pinch points, half-moons, and ``stars''} 
\label{sec:ssf}

We focus our attention on the magnetic correlations in the disordered cooperative paramagnetic regime. This is why we shall not go into the details of the low-temperature ordered phases, with the exception of the high-symmetry point $J=1/2$ on pyrochlore in Section \ref{sec:highsym}, whose nature is particularly  enlightening. This approach presents the advantage that, in the cooperative paramagnets, the properties of the kagome and pyrochlore lattices are qualitatively very similar, allowing for a parallel analysis of the two lattices.

\begin{figure}[t]
\begin{center}
\hspace{-0.5cm}\includegraphics[bb = 0 0 652 328,width=\linewidth]{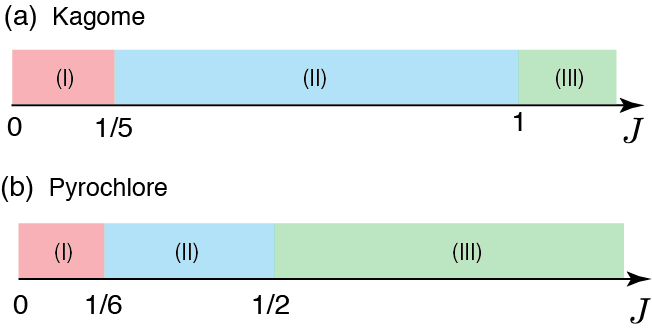}\\
\vspace{-10pt}
\caption{Ground-state phase diagram within the large-$N$ approximation. Region (I) represents the Coulomb phase where the static structure factor shows pinch points for $0< J < J_{1c}=1/5$ (kagome) and $1/6$ (pyrochlore). For $J>J_{1c}$, the flat bands do not correspond to ground states anymore, and the physics is dominated by the energy minima of the dispersive band (Figs.~\ref{fig:band_kagome} and \ref{fig:band_py}), giving rise to half-moon patterns in the static structure factor (Fig.~\ref{fig:SSF}). The high-symmetry point at $J_{2c}=1$ (kagome) and $1/2$ (pyrochlore) separates the large-$J$ region into two parts, with qualitatively different positions of the energy minima in Fourier space (Figs.~\ref{fig:band_kagome} and \ref{fig:band_py}). In the structure factor, the half-moons evolve continuously into ``star'' patterns within region (II). The boundaries have been confirmed by Monte Carlo simulations.
}
\label{fig:PD}
\end{center}
\end{figure}

\begin{figure*}[t]
\begin{center}
\hspace{0cm}\includegraphics[bb = 0 0 774 503,width=0.78\linewidth]{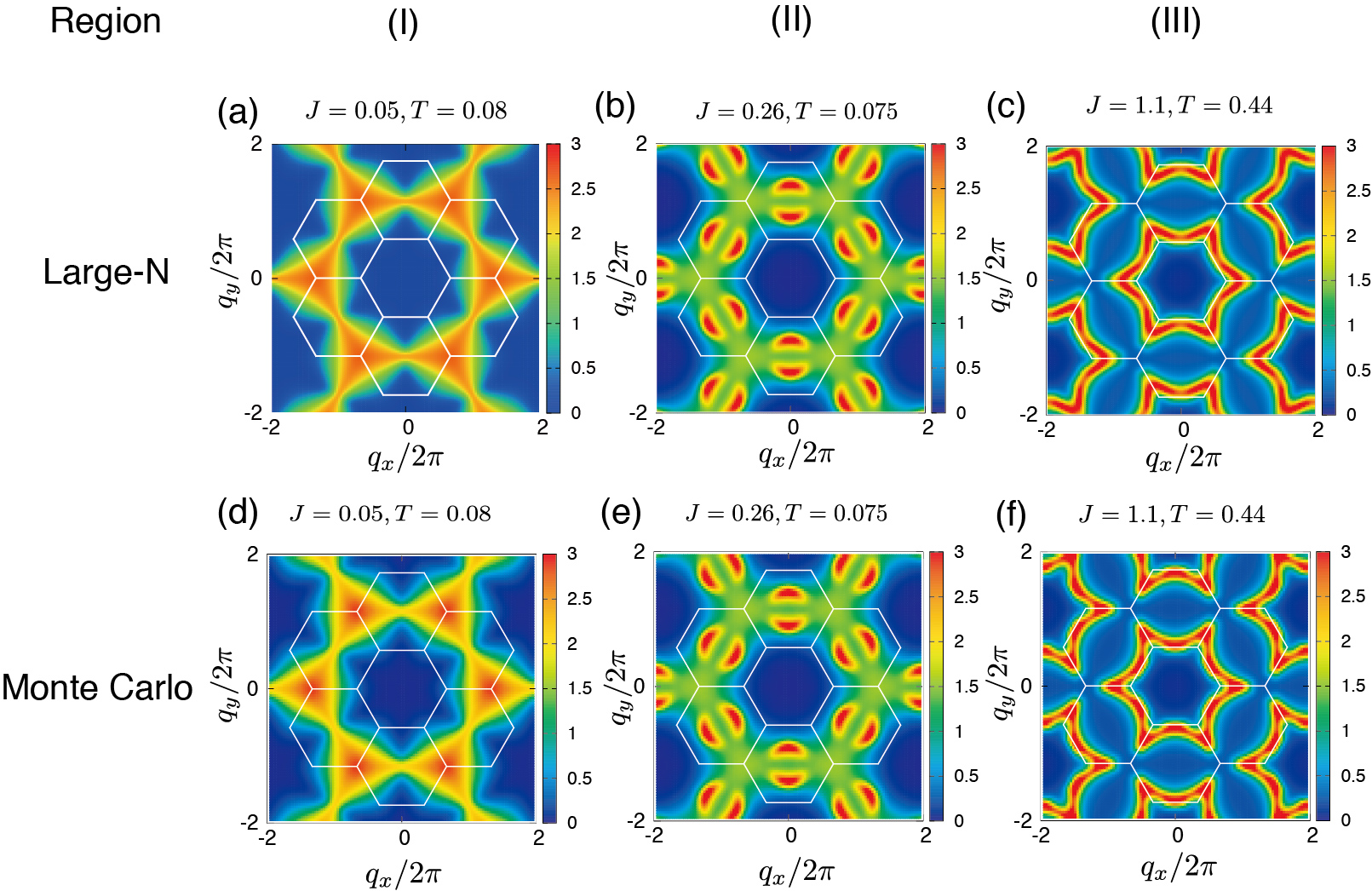}\\
-----------------------------------------------------------------------------------------------------------------------------------------\\
\hspace{0cm}\includegraphics[bb = 0 0 778 493,width=0.78\linewidth]{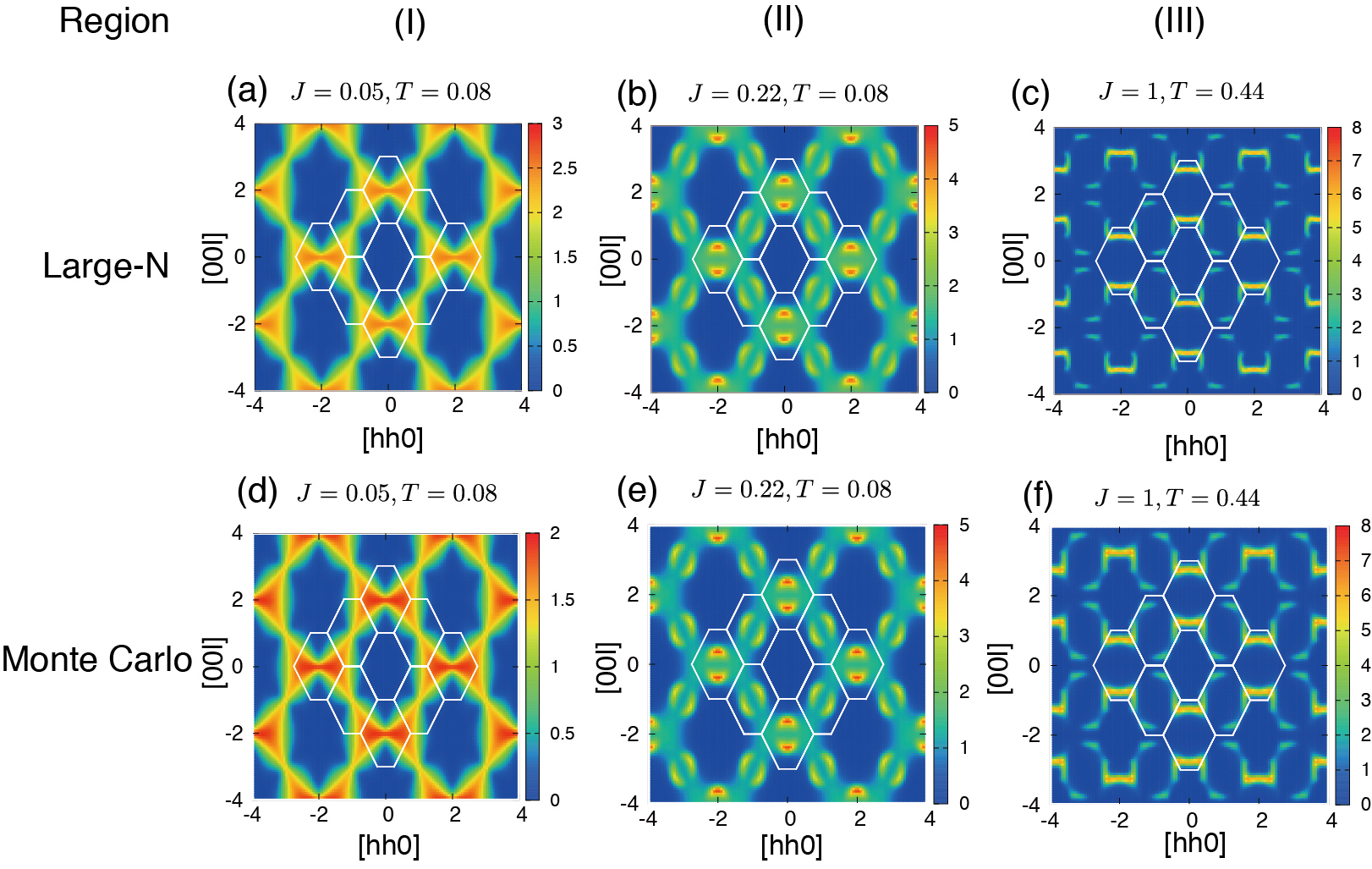}\\
\vspace{-10pt}
\caption{Structure factor for kagome (top) and pyrochlore (bottom) lattices calculated by (a)-(c) large-$N$ approximation and (d)-(f) Monte Carlo simulation. White lines denote the Brillouin zones. The model being antiferromagnetic, all the characteristic features of the scattering appear in the secondary Brillouin zone boundaries. For region (I), the pinch points are clear signatures of the divergence-free condition of the Coulomb phase. Their absence in regions (II) and (III) indicates that the system is out of the Coulomb phase. The complementary patterns of the pinch points are the half-moons (II), which adiabatically evolve into ``star'' shapes. The star patterns appear in region (II) and persist in (III). See Fig.~\ref{fig:PD} for the boundaries of the three regions.}
\label{fig:SSF}
\end{center}
\end{figure*}

The evolution of the correlations in the cooperative paramagnetic regimes are closely linked to the qualitative changes in the band structure obtained by large-$N$ analysis.
The discussion in this section relies heavily on the analysis of the low-energy band structure, supported by Monte Carlo simulations at finite temperatures. The outline of the ground-state phase diagram is given in Fig.~\ref{fig:PD} while the excellent agreement between analytics and numerics is illustrated in Fig.~\ref{fig:SSF}.

To briefly introduce the overall structure of phase diagram (Fig.~\ref{fig:PD}), the small-$J$ region, which we call region (I), is characterized by the pinch points in the structure factor [Figs. \ref{fig:SSF}(a) and \ref{fig:SSF}(d)]. Upon increasing $J$, the structure factor shows qualitative changes twice. At intermediate values of $J$, at the beginning of region (II), the structure factor develops a characteristic pattern, which we call ``half-moon" after its shape [Figs. \ref{fig:SSF}(b) and \ref{fig:SSF}(e)]. Further increasing $J$, moving continuously from region (II) to (III), $\mathcal{S}({\bm{q}})$ shows further change into the ``star" pattern. Below, we will introduce the nature of each region, separately.

\begin{figure*}[t]
\begin{center}
\hspace{0cm}\includegraphics[bb = 0 0 1174 397,width=0.9\linewidth]{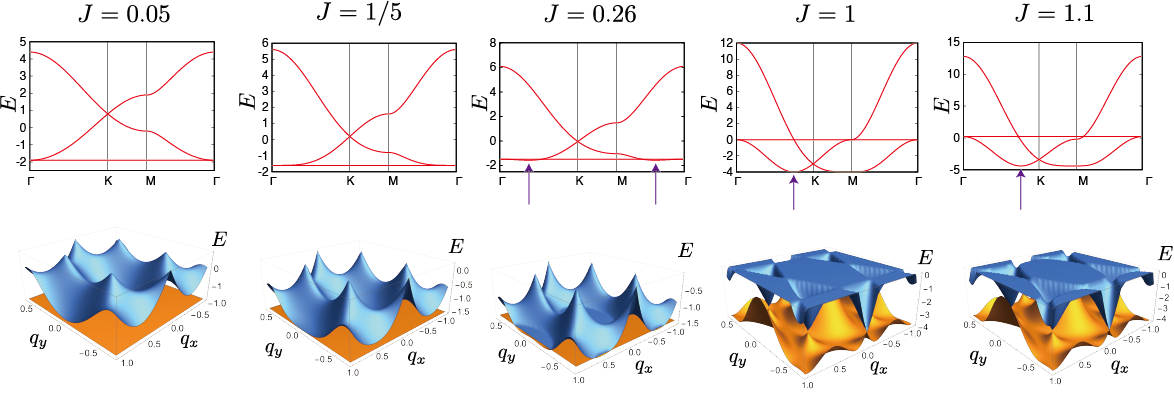}\\
\vspace{-10pt}
\caption{Kagome: band structures of $H(\bm{q})$ on high-symmetry lines (the first row), and in two-dimensional momentum space (the second row)
for several values of $J$.
In the second row the highest band is omitted for clarity. The position of energy minima are shown with black arrows.
}
\label{fig:band_kagome}
\end{center}
\end{figure*}
\begin{figure*}[t]
\begin{center}
\hspace{0cm}\includegraphics[bb = 0 0 1276 414,width=0.9\linewidth]{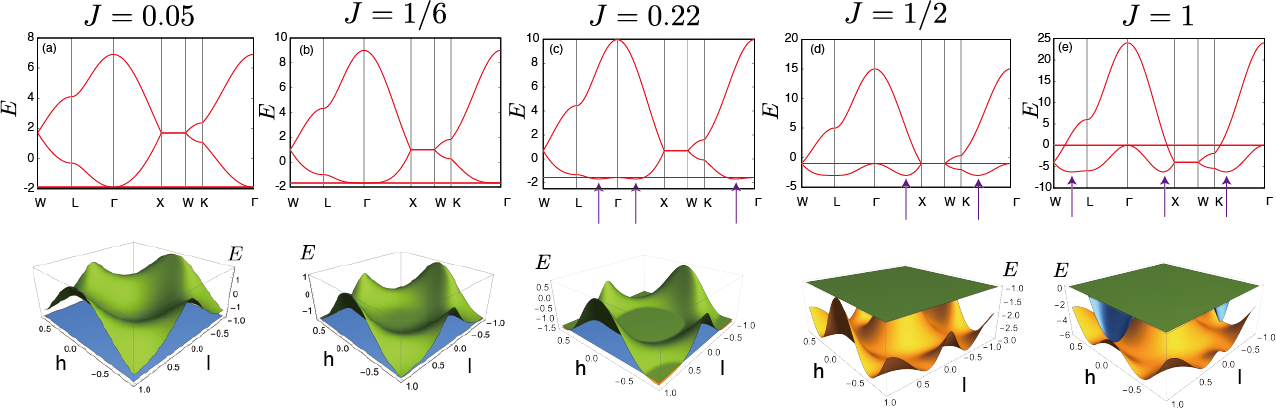}\\
\vspace{-10pt}
\caption{
Pyrochlore: band structures of $H(\bm{q})$ on high-symmetry lines (the first row), and in the $hhl$-plane (the second row)
for several values of $J$. 
In the second row the highest band is omitted for clarity. The position of energy minima are shown with black arrows.
 }
\label{fig:band_py}
\end{center}
\end{figure*}

\begin{figure}[b]
\begin{center}
\hspace{0cm}\includegraphics[bb = 0 0 424 181,width=\linewidth]{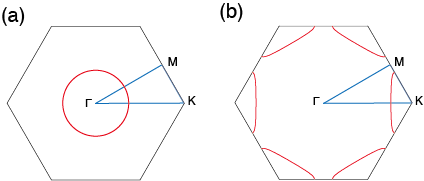}\\
\vspace{-10pt}
\caption{The energy minima in the Brillouin zone for (a) $J=0.26$ and (b) $J=1.1$ in the kagome system. 
}
\label{fig:minima_kagome}
\end{center}
\end{figure}

\begin{figure}[b]
\begin{center}
\hspace{0cm}\includegraphics[bb = 0 0 354 193,width=\linewidth]{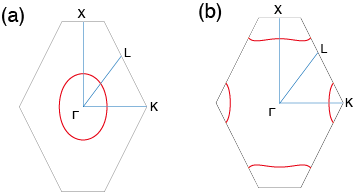}\\
\vspace{-10pt}
\caption{The energy minima in the Brillouin zone 
on a hhl-plane for (a) $J=0.22$ and (b) $J=1$ in the pyrochlore system. 
}
\label{fig:minima_pyrochlore}
\end{center}
\end{figure}
\subsection{Flat bands of the Coulomb phase}

The first noticeable outcome of the large-$N$ theory is the persistence of the flat band(s) for all values of $J$ (Figs.~\ref{fig:band_kagome} and \ref{fig:band_py}). 
The flat band persists, as long as $J_2=J_3(=J)$ is satisfied.
These flat bands, one for kagome and two for pyrochlore, are well known from the NN model ~\cite{garanin1999,canals2001,isakov2004}. 
They represent the Coulomb spin liquid where every unit cell (triangle and tetrahedron) bears zero magnetization, $\{\mathbf{M}_n=0 \;|\;\forall n\}$, and appear in the static structure factor as pinch points. The persistence of the flat bands, and their double degeneracy for pyrochlore, are readily understandable from Eq.~(\ref{Ham_conservedspin}) since all configurations of the Coulomb phase with $\mathbf{M}_n=0$ remain degenerate in presence of the farther-neighbor coupling $J$. Mathematically, this persistence takes the form of the exchange matrix $\hat{H}(\bm{q})$ being a polynomial of the NN exchange matrix~\cite{bergman2007} 
[see Appendix \ref{sec:largeNanalytics}, in particular Eqs.~(\ref{eq:2ndhopping}) and (\ref{eq:poly_3})]. 
As such, the two exchange matrices share the same basis of eigenvectors and the flatness of eigenvalues is transmitted from the latter to the former.

Hence, one needs to consider the evolution of the dispersive bands to understand the qualitative changes in the magnetic correlations as a function of $J$.

\begin{figure}[t]
\centering\includegraphics[width=\columnwidth]{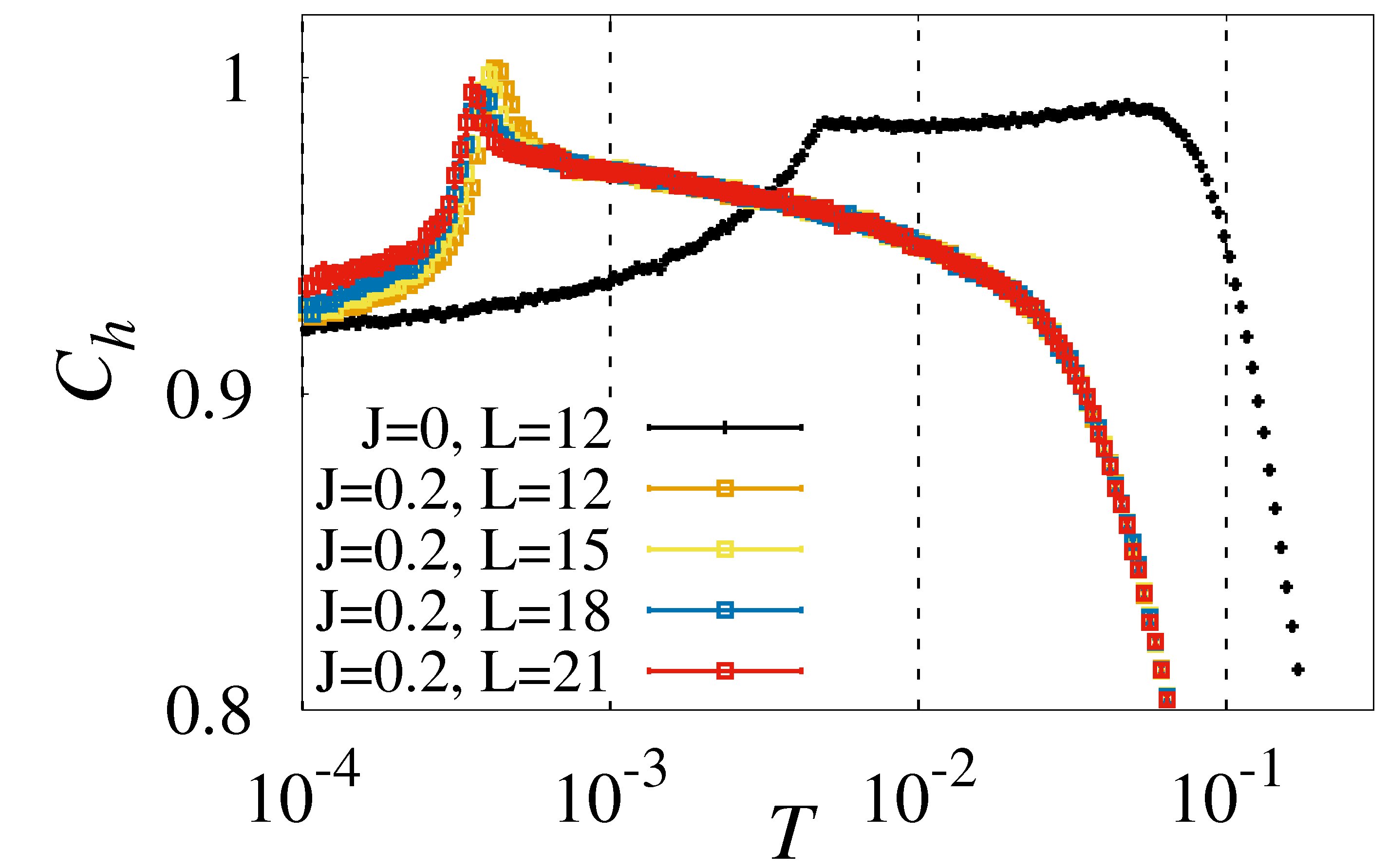}\\
\centering\includegraphics[width=\columnwidth]{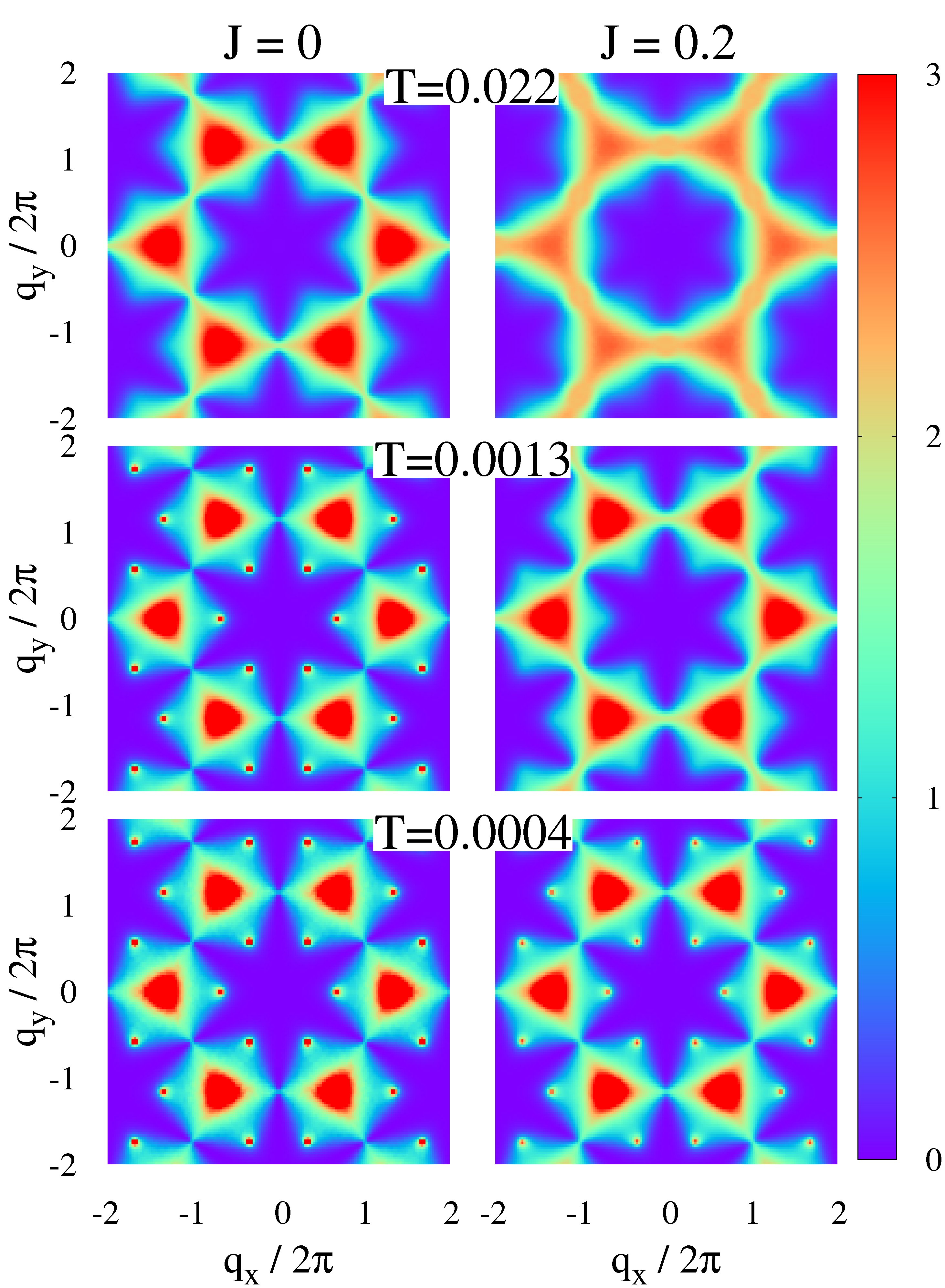}
\caption{Kagome:
Top: Heat capacity $C_{h}$ showing the low-temperature bump/kink into the coplanar regime at $T\approx 0.005$ for $J=0$ (black dots) and at $T\approx 0.0004$ for $J_{1c}=1/5$ (colored triangles).
Bottom: Temperature evolution of the static structure factor $\mathcal{S}(\mathbf{q})$ for $J=0$ (left) and $J_{1c}=1/5$ (right), obtained by the classical Monte Carlo simulation.
}
\label{fig:ChK}
\end{figure}

\begin{figure}[t]
\centering\includegraphics[width=0.8\columnwidth]{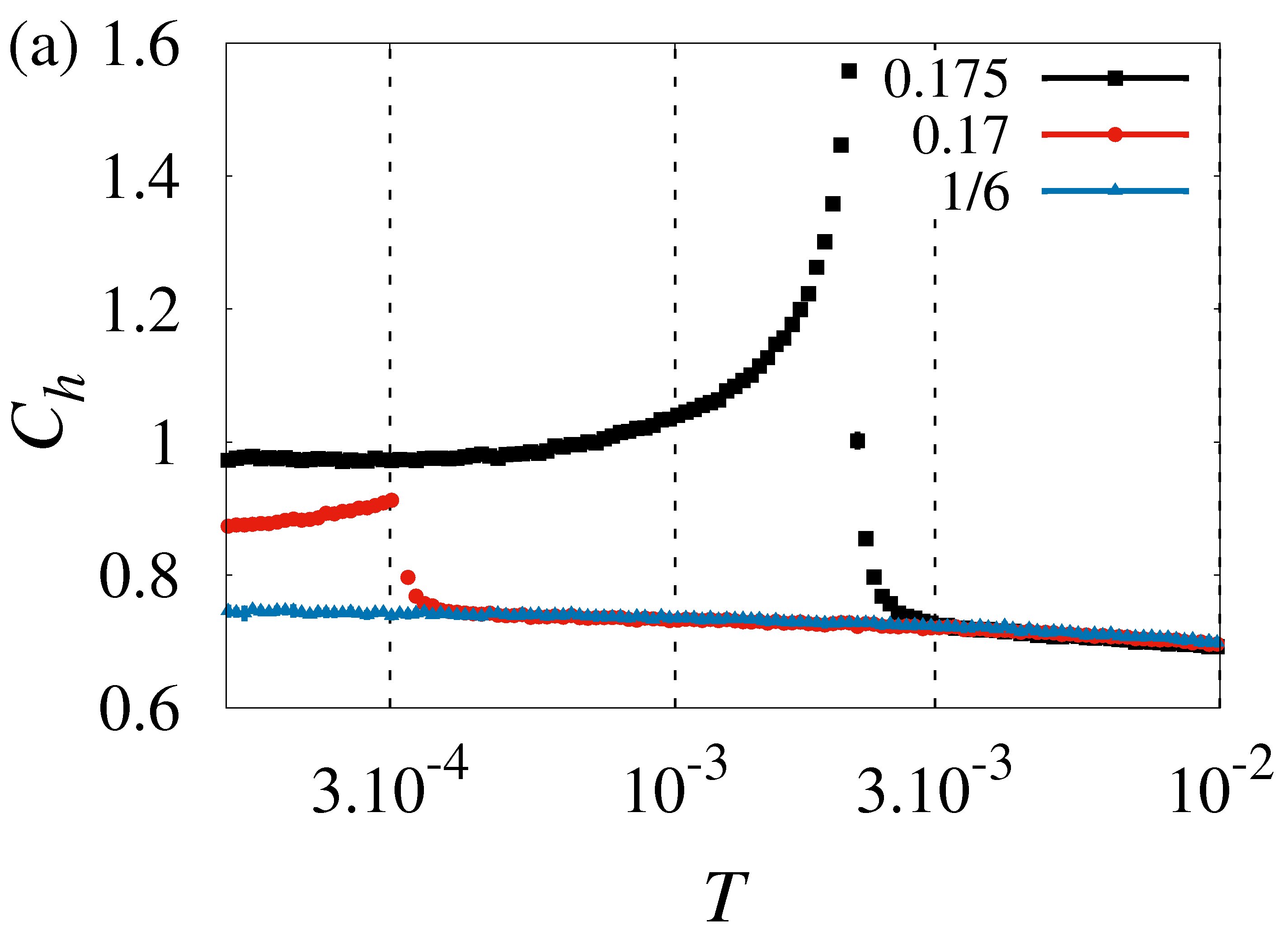}\\
\centering\includegraphics[width=\columnwidth]{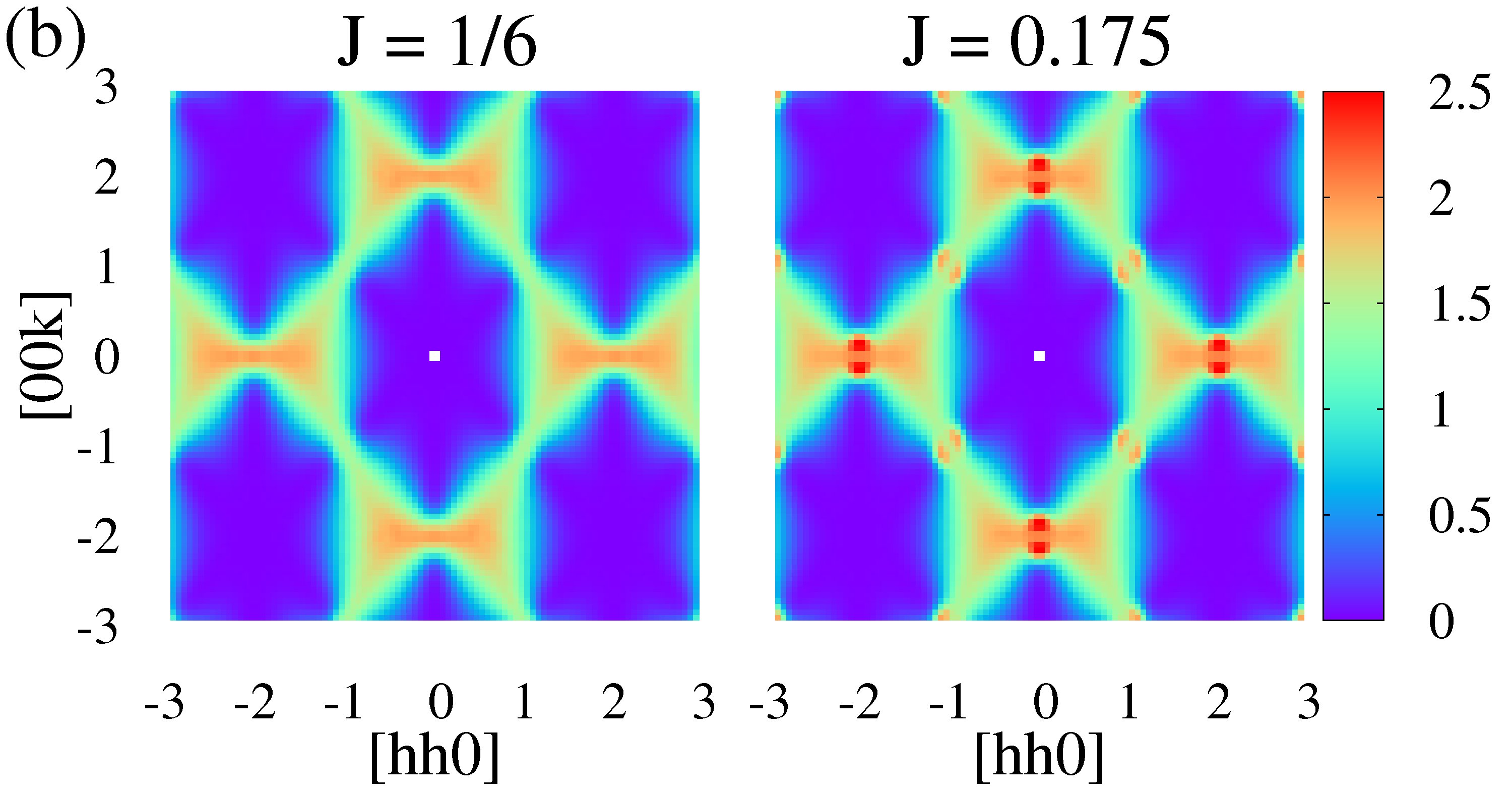}
\caption{Pyrochlore:
(a) Heat capacity $C_{h}$ showing the low-temperature ordering when $J$ is just above $J_{1c}$, but not at $J_{1c}=1/6$, down to $T\sim 10^{-4}$. 
(b) Static structure factor $\mathcal{S}(\mathbf{q})$ in the [hhl] plane at $T=0.005$ showing the pinch points at $J_{1c}$ replaced by very small half moons above $J_{1c}$.
}
\label{fig:ChSqP}
\end{figure}

\subsection{From pinch points to half-moons, near $J_{1c}$}
The flat bands form the ground-state manifold up to $J = J_{1c}=1/5$ for kagome \cite{li2018} and $1/6$ for pyrochlore. This delimits the region (I) of the phase diagram of Fig.~\ref{fig:PD}. For $J>J_{1c}$, one of the dispersive bands has a lower energy than the flat bands in parts of the Brillouin zone [Figs.~\ref{fig:band_kagome} and \ref{fig:band_py}], and the energy minima form a closed line (surface) in a Brillouin zone for a kagome (pyrochlore) lattice [Fig.~\ref{fig:minima_kagome}, \ref{fig:minima_pyrochlore}]. It means that the static structure factor is now dominated by a dispersive band rather than the flat band. As a consequence, the pinch points are smoothed out and their non-analyticality disappears, leaving behind half-moon patterns at the center of the Brillouin zone. The half-moons, and later ``star'' patterns, are characteristic of the region $J>J_{1c}$, and can be regarded as complementary to the pinch points (see discussion in Sec. \ref{sec:origin}).

The kagome NN Heisenberg antiferromagnet ($J=0$) is well known for its Coulomb phase at intermediate temperatures, followed by a coplanar regime at lower temperatures selected by thermal order by disorder~\cite{chalker1992,huse1992,reimers1993,zhitomirsky2008}. 
The Coulomb phase is marked by a plateau in the heat capacity and pinch points in the structure factor. 
When the system enters the coplanar regime, the heat capacity exhibits a kink, and sharp peaks of scattering at $\mathbf{q}_{\sqrt{3}}=(4\pi/3,0)$ appear in the structure factor (Fig.~\ref{fig:ChK}). 
These peaks represent the onset of the $\sqrt{3}\times\sqrt{3}$ long-range order as $T\rightarrow 0^{+}$ [\onlinecite{chern2013}], but are not Bragg peaks since there is no dipolar long-range order at finite temperature.

At $J=J_{1c}$, 
the softening of the 
band touching between the lowest dispersive and the flat bands enhances thermal fluctuations compared to $J=0$. As a consequence, the coplanar regime is pushed down to lower temperatures by an order of magnitude (Fig.~\ref{fig:ChK}). Noticeably, at intermediate temperatures ($T=0.022$), the pinch points visible at $J=0$ have disappeared in favor of the onset of the characteristic half-moons for $J=J_{1c}$. This is why the heat capacity does not show the characteristic plateau of the Coulomb phase at $J_{1c}$.\\

As for the pyrochlore lattice, the value of $J_{1c}=1/6$ obtained from large-$N$ is confirmed by simulations down  to $T \sim 10^{-4}$ (Fig.~\ref{fig:ChSqP}). For $J=J_{1c}^{+}$, just above the boundary inside region (II), the system orders, but it remains disordered at the boundary $J_{1c}$. As a consequence, the pinch points of the Coulomb phase are visible up to $J=J_{1c}$, replaced by half-moons as soon as the system enters region (II) [Fig.~\ref{fig:ChSqP}.(b)]. Please note that the small thickness of the pinch points for $J=J_{1c}$ is due to the proximity of the half-moon regime at finite temperature.

\subsection{Origin of the half-moons \label{sec:origin}}

In region (II), the structure factor develops half-moon patterns [Fig.~\ref{fig:SSF}(b) and \ref{fig:SSF}(e)].
What kind of information can be read from this characteristic magnetic scattering? The intensity of scattering at particular wave vectors $\bm{q}$ is determined by the energy of magnetic modes and the corresponding real-space structure of dominant modes. In our large-$N$ language, they are described by the shape of the energy-minima manifold, $\varepsilon_{\rm min}(\bm{q})$, and the weight of the corresponding eigenfunctions, $\Xi_\eta(\bm{q}) \equiv \sum_{\mu, \nu}  [\bm{\psi}^\ast_\eta (\bm{q})]_\mu [\bm{\psi}_\eta (\bm{q})]_\nu $. The weight, $\Xi_\eta(\bm{q})$, satisfies the sum rule:  
\begin{eqnarray}
\sum_{\eta}  \Xi_\eta(\bm{q})= N_{\rm sub}, 
\label{eq:sum_rule}
\end{eqnarray}
due to the normalization of spin length.

\begin{figure}[b]
\begin{center}
\includegraphics[bb = 0 0 694 250,width=\linewidth]{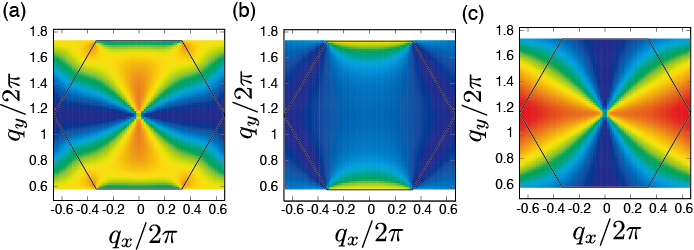}\\
\vspace{-10pt}
\caption{\textit{Kagome:} The weights of eigenvectors, $\Xi_\eta(\bm{q})$, in the second Brillouin zone centered at $\left(0, \frac{4\pi}{\sqrt{3}} \right)$, for (a) a dispersive band with lower energy, (b) a dispersive band with higher energy, and (c) a flat band.
Note the complementary singularities at the zone center in the left and right panels. 
}
\label{fig:ev_kagome}
\end{center}
\end{figure}

Basically, the high-intensity points of the half-moons follow the position of the energy minima. However, the energy minima do not account for everything. On one hand, the energy-minima manifold, $\varepsilon_{\rm min}(\bm{q})$, is extended in Fourier space. More precisely, it can be defined locally as a hypersurface; a closed line for the two-dimensional kagome and a closed surface for the three-dimensional pyrochlore (see Figs.~\ref{fig:minima_kagome} and \ref{fig:minima_pyrochlore}, and {Appendices~\ref{sec:appendix_kagome} and \ref{sec:appendix_pyrochlore}} for more details). On the other hand, it is clear from Figs.~\ref{fig:SSF}(b) and \ref{fig:SSF}(e) that the extension of the half-moons is finite. They terminate at some point and do not form closed curves as expected from the energy-minima manifold. This vanishing intensity was coined as ``ghost'' excitations for the kagome Heisenberg antiferromagnet \cite{robert2008}.

\begin{figure}[b]
\centering\includegraphics[width=0.45 \columnwidth]{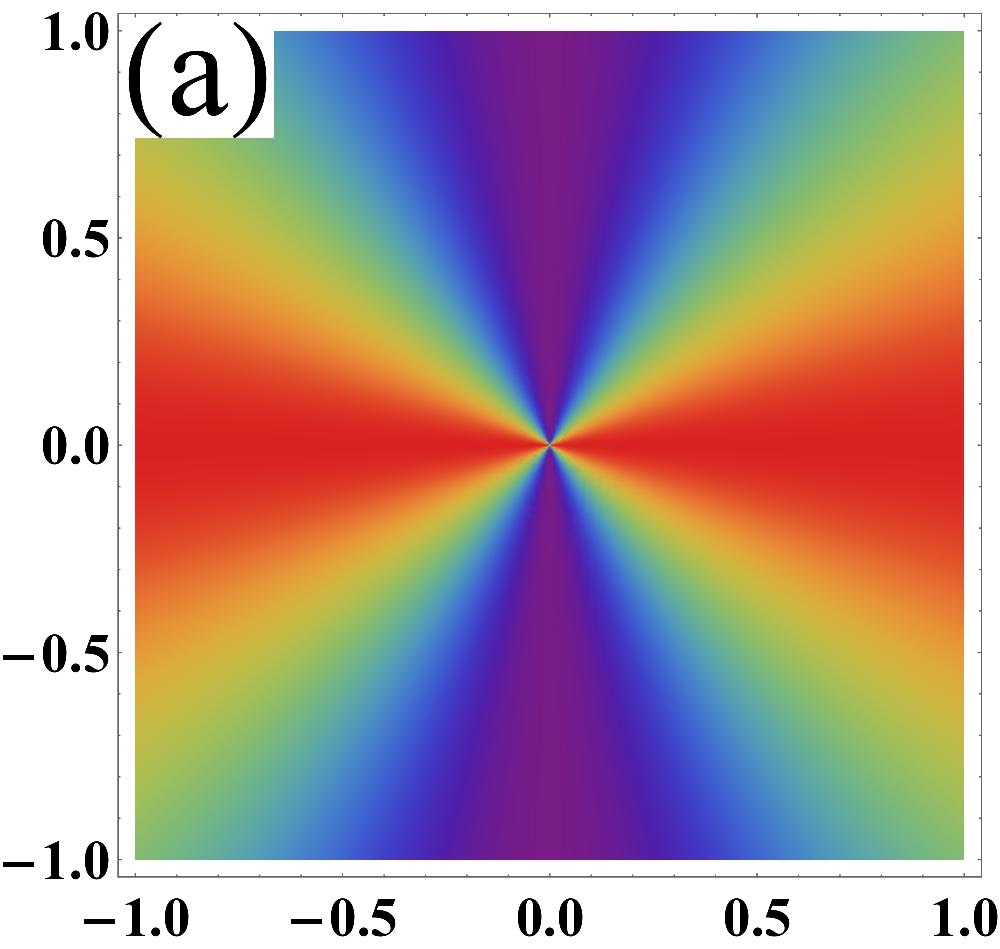}
\centering\includegraphics[width=0.45 \columnwidth]{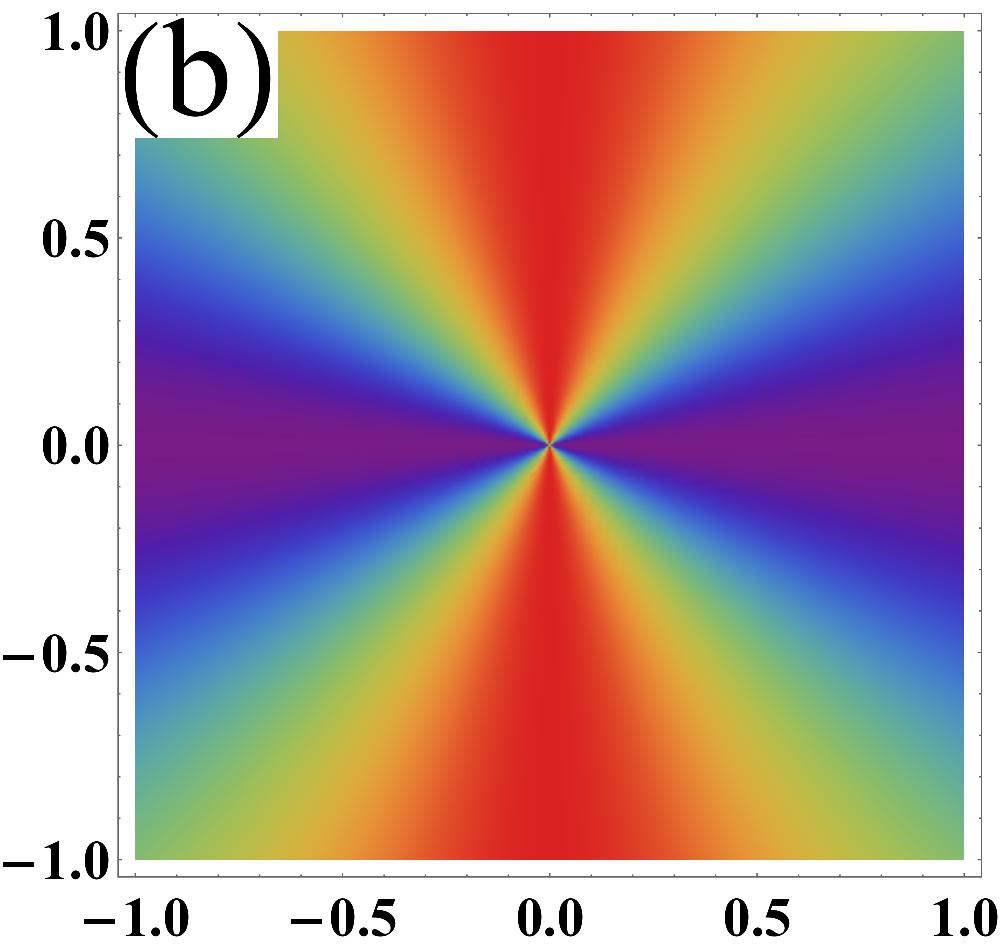}\\
\centering\includegraphics[width=0.45 \columnwidth]{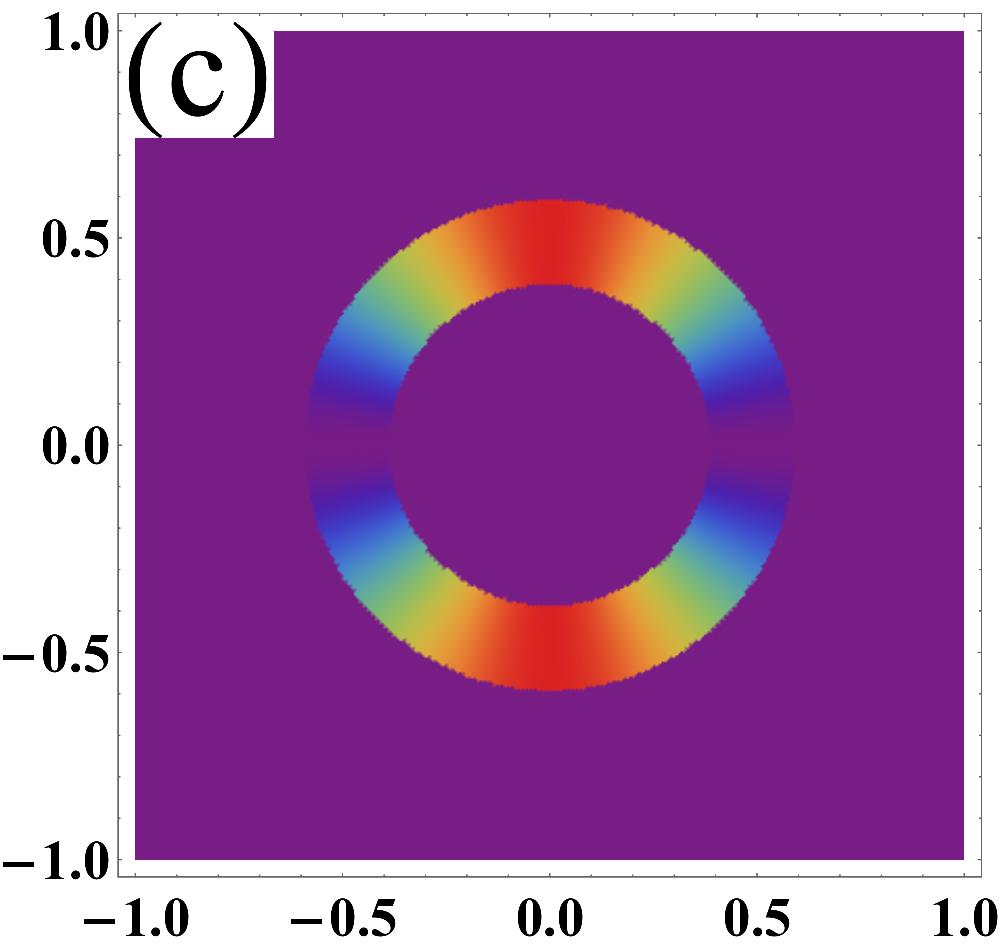}
\centering\includegraphics[width=0.45 \columnwidth]{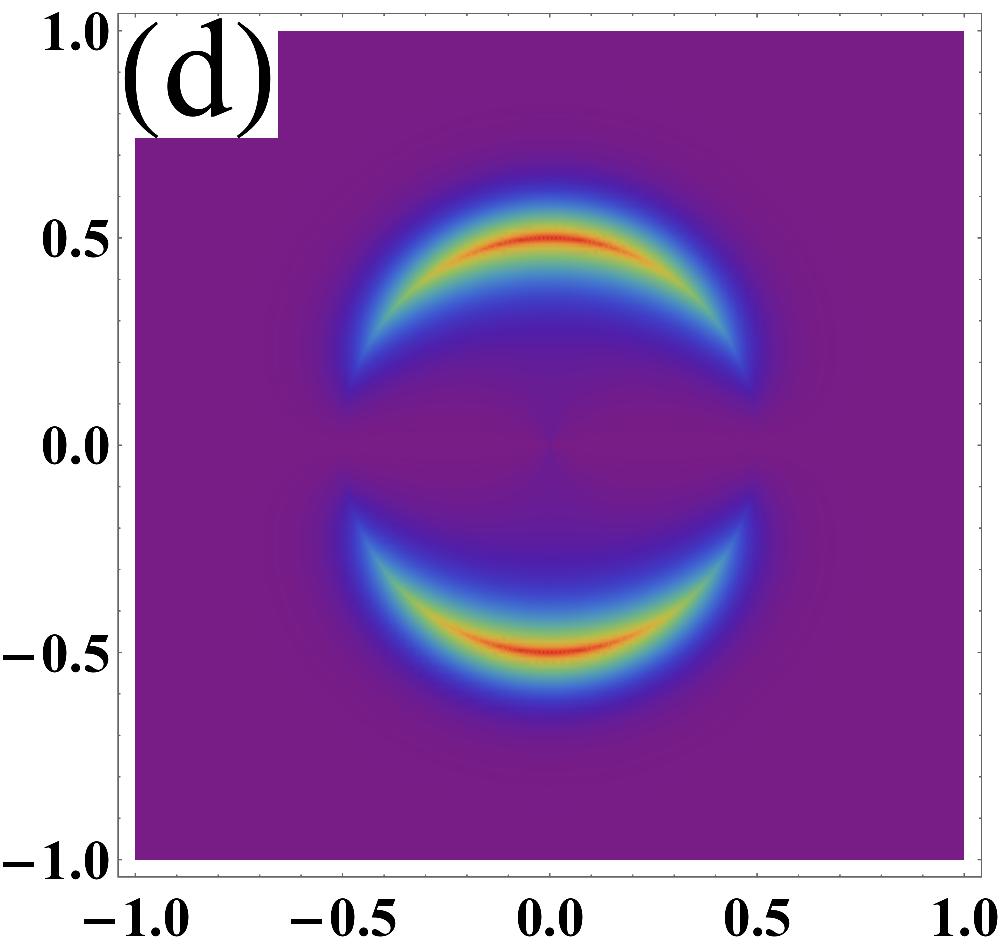}
\caption{
Schematic illustration of the emergence of half-moons at the center of a Brillouin zone (in arbitrary units). (a) The pinch point of the flat band. (b) Its complement. (c) An annular cut-out of the pinch-point complement, around the energy minima $\varepsilon_{\rm min}(\bm{q})$. (d) The pinch-point complement multiplied by a Gaussian of radius delimited by $\varepsilon_{\rm min}(\bm{q})$.
}
\label{fig:illus}
\end{figure}

This discrepancy, symbolized by a missing arc that should connect the half-moon pair, can be attributed to the spatial character of the magnetic mode. In Fig. \ref{fig:ev_kagome}, we show the intensity maps of $\Xi_\eta (\bm{q})$ for a kagome lattice in the second Brillouin zone, surrounding the wave vector $\left(0, \frac{4\pi}{\sqrt{3}} \right)$. The traditional pinch point resides in the flat band [Fig. \ref{fig:ev_kagome}(c)]. Half-moons are from the lower-dispersive band [Fig. \ref{fig:ev_kagome}(a)]. From $\Xi_\eta(\bm{q})$, one finds that the high-intensity regions of flat mode and lower-dispersive mode complement each other. 
This complementarity is originated in the sum rule, Eq.~(\ref{eq:sum_rule}).
Around the pinch point, the contribution from the highest-energy band is small, and the sum rule is satisfied only between the flat mode and the lower dispersive mode.
The missing arc is attributed to the vanishing weight of lower dispersive band in the bow-tie region, where the flat band contribution is dominant.
In this sense, the missing arc of the half-moon can be considered as a shadow of pinch point, thus answering the 10-year old open question about the nature of the ``ghost'' excitations in the kagome Heisenberg antiferromagnet \cite{robert2008}. This missing arc signals the proximate presence of a pinch point, and serves as evidence that the system is in the vicinity of a Coulomb phase. 

The half-moon formation can be discussed in a more general context not specific to the kagome system.
Given the pinch point reflects a singularity of the flat band eigenvectors as a function of momentum, and the completeness of the
eigenvector basis, there must be a complementary non-analyticity in (at least) one of the other bands [Figs.~\ref{fig:illus} (a) and \ref{fig:illus} (b)]. As these are in general not flat,
their constant energy cuts at small radius [Ref.~\onlinecite{note_radius}] will typically yield the shape of an annulus radially, with an
angular modulation characteristic of the longitudinal pinch-point projector [Figs.~\ref{fig:illus}(c) and \ref{fig:illus}(d)]. These combine to yield (a pair of) half moons. Depending on
the relative ordering of the bands in energy, these half moon pairs may either appear in the ground state correlations (Fig.~\ref{fig:SSF}) or in the excitation spectrum (see Sec.~\ref{sec:semiclassical_dynamics}).

\begin{figure}[ht]
\begin{center}
\includegraphics[bb = 0 0 696 513,width=\linewidth]{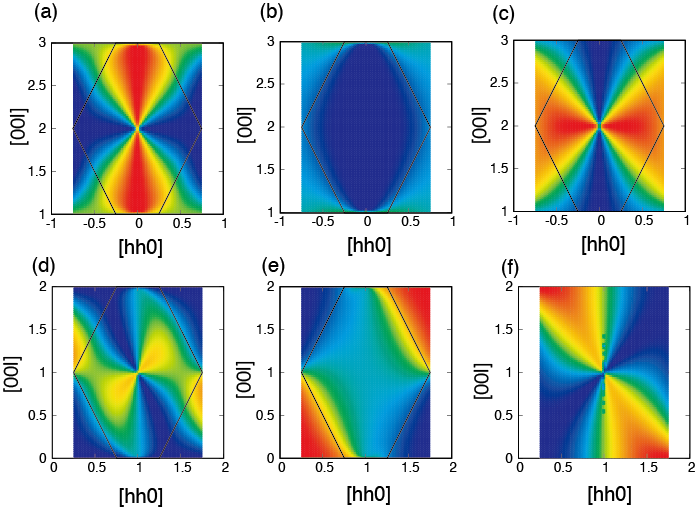}\\
\vspace{-10pt}
\caption{Pyrochlore: the weights of eigenvectors, $\Xi_\eta(\bm{q}) $, in the second Brillouin zone centered at $\left[0 0 2 \right]$ for (a)-(c) and at $\left[1 1 1 \right]$ for (d)-(f). (a) and (d) are for a dispersive band with lower energy, (b) and (e) for a dispersive band with higher energy, and (c) and (f) for the summed contribution of the two flat bands.
}
\label{fig:ev_pyrochlore}
\end{center}
\end{figure}

The same scenario holds for the pyrochlore lattice: the maps are shown in 
Figs. \ref{fig:ev_pyrochlore}(a)-\ref{fig:ev_pyrochlore}(c) (centered at $\left[002\right]$) and \ref{fig:ev_pyrochlore}(d)-\ref{fig:ev_pyrochlore}(f) (centered at $\left[111\right]$). 
Again, the combination of the energy-minima surface and the intensity map $\Xi_\eta(\bm{q})$ gives rise to the half-moon patterns.

This explanation remains valid throughout the phase diagram for $J>J_{1c}$, and in particular as the half-moons continuously deform into star patterns.

\subsection{From half-moons to star patterns for $J > J_{1c}$}
\label{sec:HMS}

\begin{figure}[ht]
\centering\includegraphics[width=0.97\columnwidth]{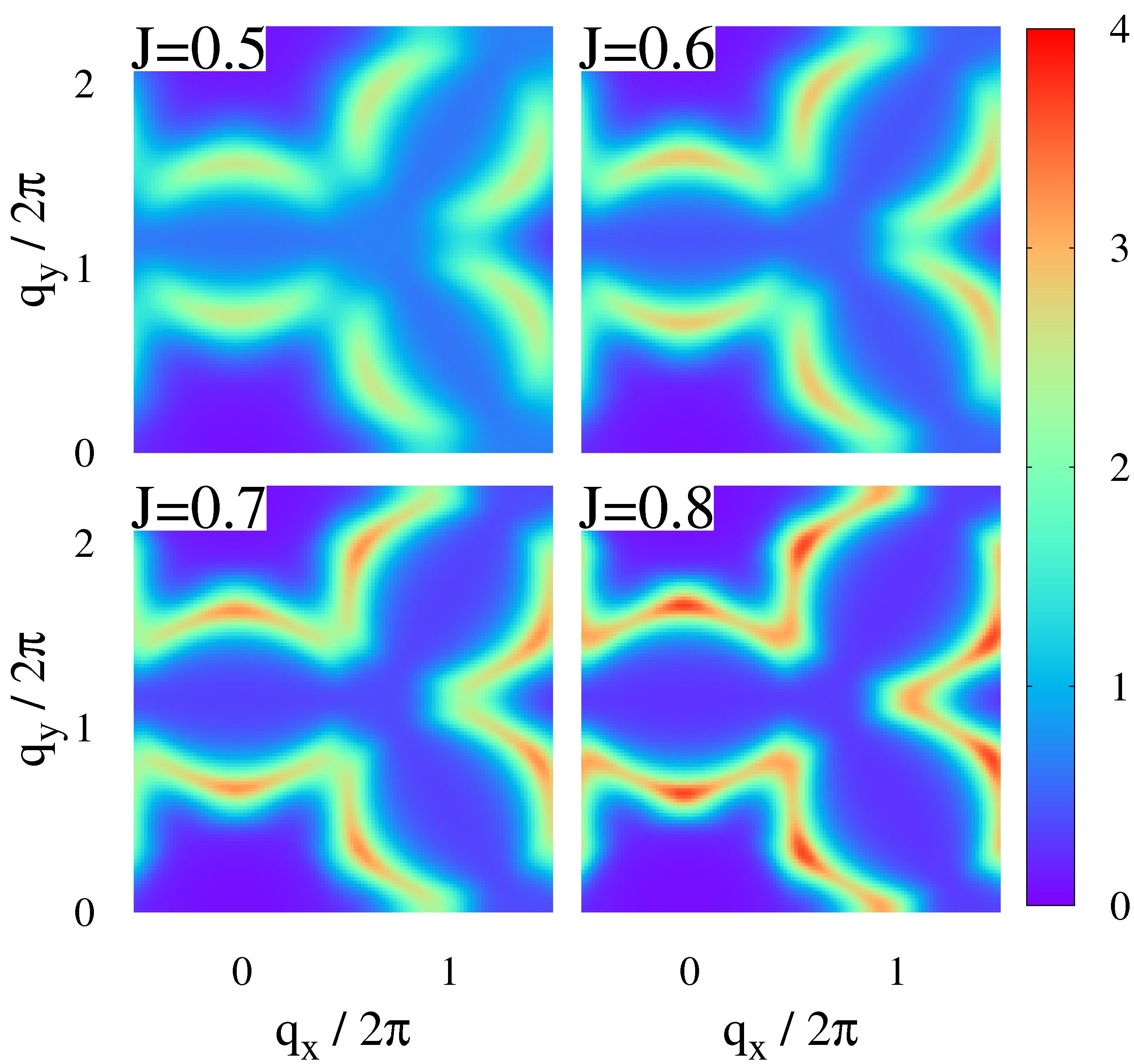}
\caption{Kagome:
evolution from half-moon to star patterns in the static structure factor $\mathcal{S}(\mathbf{q})$ for $J_{1c}<J<J_{2c}$, obtained by Monte Carlo simulation. The simulation temperature is $T=0.32$.
}
\label{fig:HMK}
\end{figure}

\begin{figure}[ht]
\centering\includegraphics[width=0.97\columnwidth]{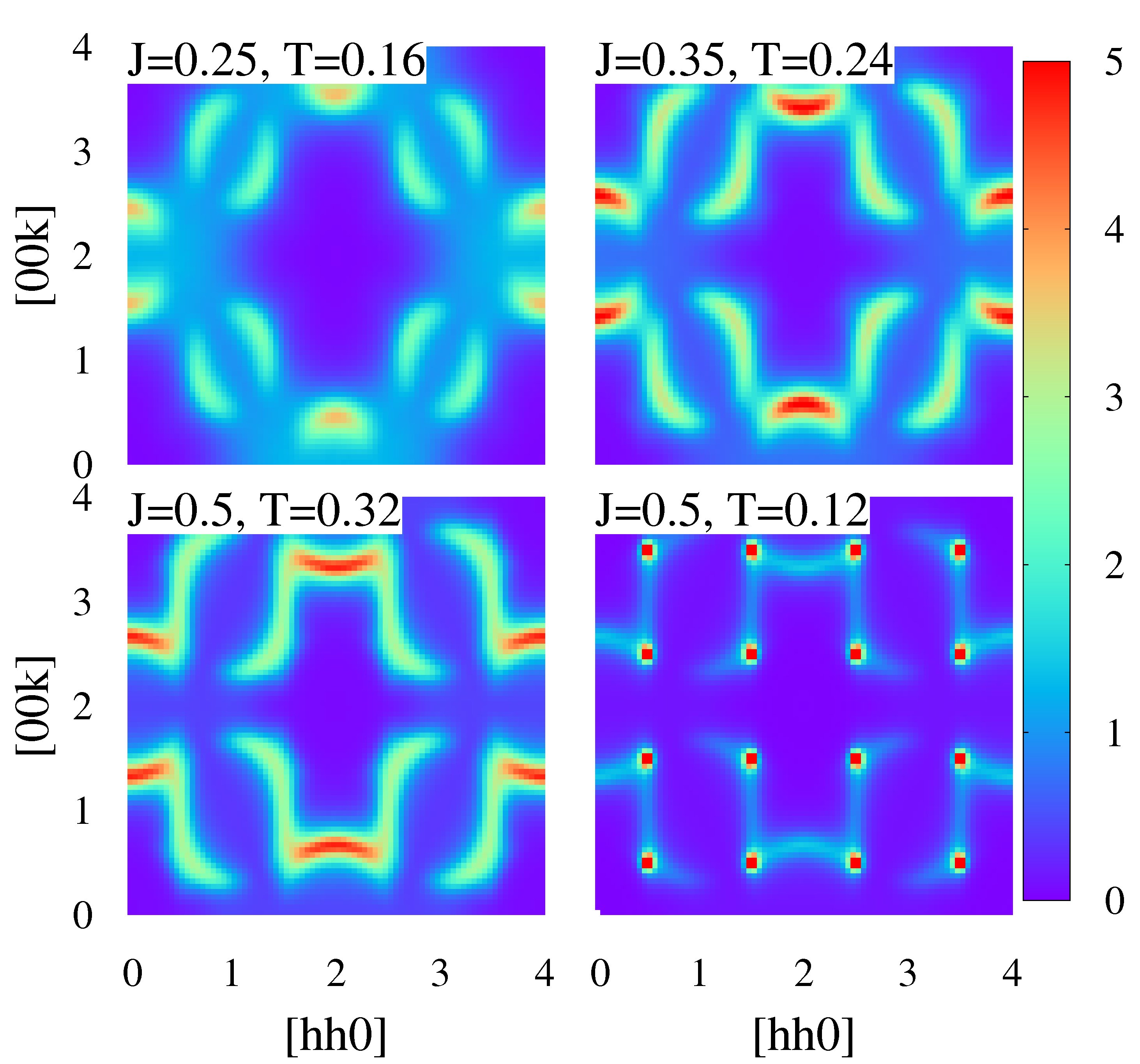}
\caption{Pyrochlore:
evolution from half-moon to star patterns in the static structure factor $\mathcal{S}(\mathbf{q})$ for $J_{1c}<J<J_{2c}$, obtained by Monte Carlo simulation. The bottom right panel is below the transition temperature at the high-symmetry point $J=J_{2c}$ with Bragg peaks at $\mathbf{q}_{L}=\left[\frac{1}{2}\frac{1}{2}\frac{1}{2}\right]$.
}
\label{fig:HMP}
\end{figure}

In the previous subsection, we have seen how the shape of the half-moon is linked to the position of the energy minima in Fourier space. These energy minima continuously move as a function of $J$, and the shape of the half-moon evolves with them, as illustrated in Figs.~\ref{fig:HMK} and \ref{fig:HMP}.


Upon increasing $J$, the radius of the half-moons increases. Since the radius is limited by the size of the Brillouin zone, neighboring half-moons eventually connect to each other by their extremities, forming star shapes. Please note that while the star shapes are rather obvious for kagome [see the top panels of Figs.~\ref{fig:SSF}.(c) and \ref{fig:SSF}(f) and Fig.~\ref{fig:HMK}], they are somewhat more figurative for pyrochlore in a [hhl] plane [see the bottom panels of Figs.~\ref{fig:SSF}.(c) amd \ref{fig:SSF}(f) and Fig.~\ref{fig:HMP}]. For convenience, we shall use the name of ``star'' for both lattices, which shall be understood as the patterns formed by connected half-moons.

\begin{figure}[ht]
\centering\includegraphics[width=0.8\columnwidth]{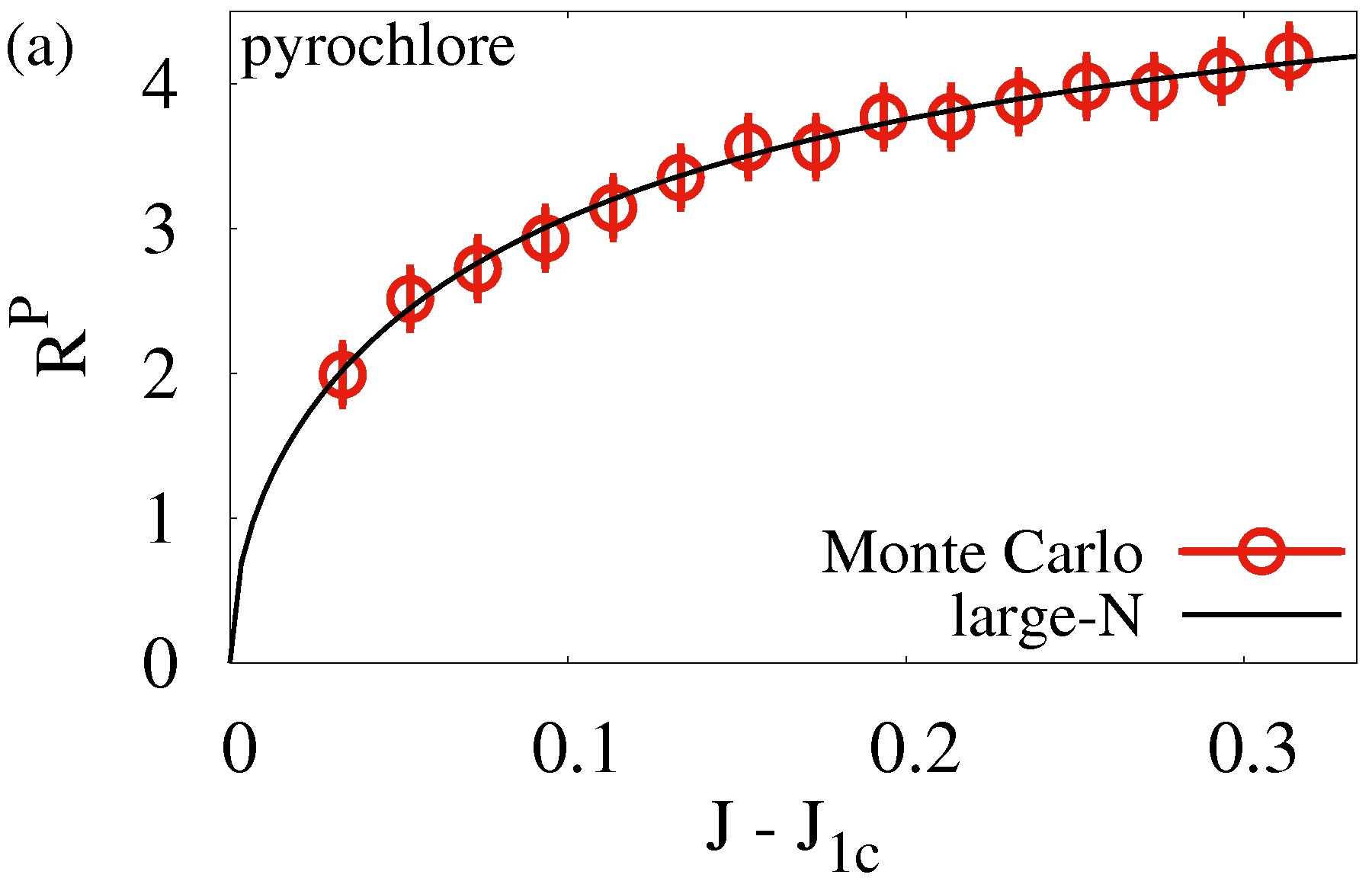}
\centering\includegraphics[width=0.8\columnwidth]{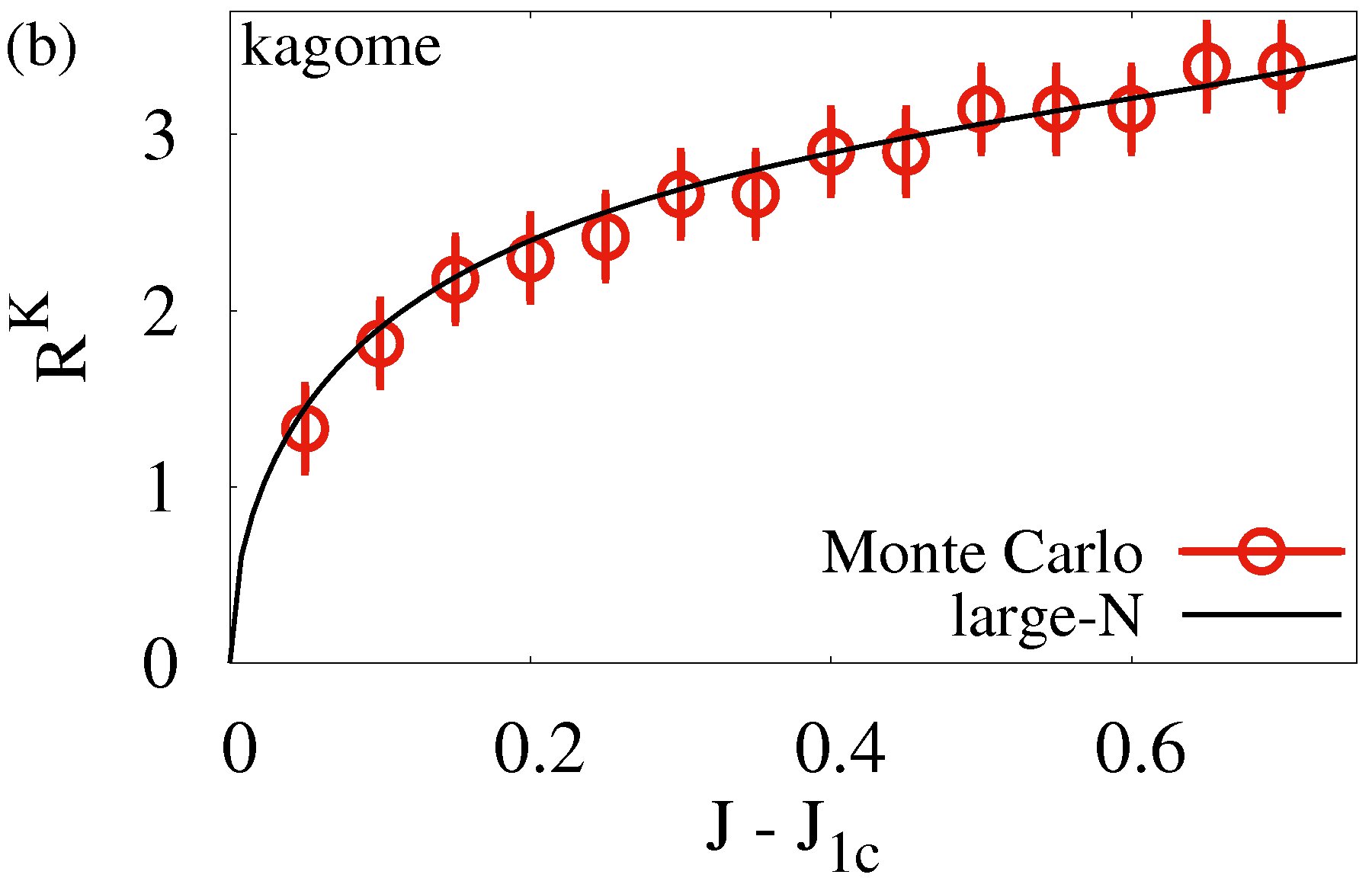}
\caption{
The evolution of the radius of the half moons in Fourier space 
(normalized by $2\pi$)
for pyrochlore (a) and kagome (b) agrees quantitatively between numerics (red circles) from classical Monte Carlo simulations and analytics (black line) derived from large-$N$ calculations [Eqs.~(\ref{eq:radiusK}) and (\ref{eq:radiusP})]. 
The error bars come from the discretization of the Fourier space in a system of finite size ($L=30$).
For kagome, data were taken at $T=0.32$. For pyrochlore, data were taken for a range of temperatures above the transition temperature, $T\in[0.08;0.24]$.
}
\label{fig:radius}
\end{figure}

The increase of the half-moon radius in Fourier space, $R^{K,P}$, can be calculated analytically thanks to the large-$N$ method
\begin{align}
R^{\mathrm{K}} = \frac{4}{\sqrt{3}} \arccos \sqrt{\frac{1}{8} \left[ \left(\frac{1+J}{2J} \right)^2 -1 \right] },
\label{eq:radiusK}
\end{align}
along the $\Gamma$M direction for kagome, and 
\begin{align}
R^{\mathrm{P}} =  2  \arccos  \left[ \frac{4J + 1 - 28J^2 }{32J^2} \right], 
\label{eq:radiusP}
\end{align}
along the $\Gamma$X direction for pyrochlore. 
These formulas are in excellent agreement with results obtained from Monte Carlo simulations in the regime of collective paramagnetism, and above any potential transition temperature (Fig.~\ref{fig:radius}).

\section{Semi-classical dynamics}
\label{sec:semiclassical_dynamics}

In experiments, signals of anomalous magnetic correlation are sometimes observed in finite-frequency regions, through, e.g., inelastic neutron scattering. For example, it is at finite energy of the kagome Heisenberg antiferromagnet that half-moons were first observed \cite{robert2008}, before being stabilized as signature of the ground state, at low energy, via farther-neighbor interactions~\cite{udagawa2016,rau2016,mizoguchi2017}. Accordingly, in order to find the half-moons and stars in a realistic experimental setting, it is desirable to estimate the energy scale of the corresponding magnetic excitations.
In the context of the large-$N$ analysis, these magnetic patterns are associated with energy bands (Fig.s~\ref{fig:band_kagome} and \ref{fig:band_py}), but this band energy cannot be interpreted as the frequency of experimental probes in itself. A relation connecting them was proposed under the assumption of relaxational dynamics \cite{conlon2009,conlon2010}. However, it is not {\it a priori} obvious if this assumption holds in our system.
To this end, in this section, we address the dynamics of the system by solving the semiclassical LL equation in Eq. (\ref{eq:LL})
and calculating the dynamical structure factor $\mathcal{S}(\bm{q},\omega)$. Here, we focus on regions (II) and (III) for both kagome and pyrochlore lattices. 

\begin{figure}[b]
\begin{center}
\includegraphics[bb = 0 0 339 332,width=\linewidth]{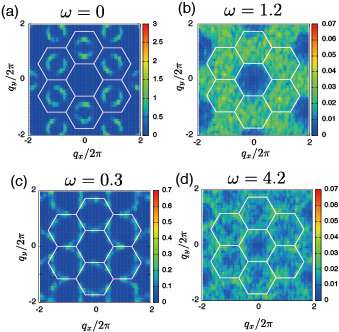}\\
\vspace{-10pt}
\caption{Kagome:
$\omega$-cuts of $S(\bm{q}, \omega)$
for (a), (b) $J = 0.3, T= 0.05$ and (c), (d) $J = 1.1, T= 0.275$. 
At low frequency, the half-moon/star shape are clearly visible.
}
\label{fig:llg_kagome}
\end{center}
\end{figure}

Let us first see the results for a kagome lattice.
Cuts for several frequencies are shown in Figs. \ref{fig:llg_kagome}. 
As expected, the characteristic patterns observed for $\mathcal{S}({\bm{q}})$ are obtained in the low-energy sectors in both regions.
In region (II), as clearly seen in Fig.~\ref{fig:llg_kagome}(a), the half-moon signal appears in the quasi-elastic regime: $\omega=0$,
showing that this pattern dominates the long-time behavior of magnetic correlation in this region. 
The pattern sustains with small $\omega$ dependence
in the low-frequency region. Upon going to intermediate energy scale comparable to NN coupling, $\omega\sim 1$, the signal smears out.
Similarly, in region (III), the star shape appears in the low-frequency part of the dynamical structure factor [Fig.~\ref{fig:llg_kagome}(c)], and it gradually blurs towards higher energy.
The remnant of the star pattern remains observable up to higher energy, compared with the vanishing of half-moons in region (II), probably attributed to
the growing energy scale of magnetic modes as $J$, as implied by the larger bandwidth obtained in the large-$N$ analysis (Fig.s~\ref{fig:band_kagome} and \ref{fig:band_py}).

\begin{figure}[t]
\begin{center}
\includegraphics[bb = 0 0 379 373, width=\linewidth]{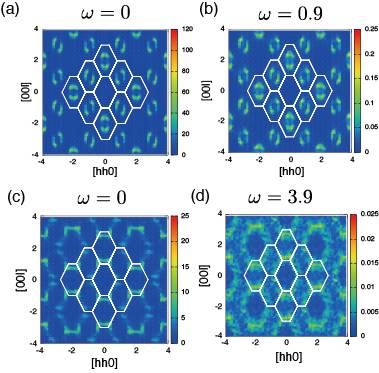}\\
\vspace{-10pt}
\caption{Pyrochlore:
$\omega$-cuts of $\mathcal{S}(\bm{q}, \omega)$
for (a), (b) $J = 0.22, T= 0.05$, and (c), (d) $J = 1, T= 0.5$.
}
\label{fig:llg_pyrochlore}
\end{center}
\end{figure}

In the Heisenberg antiferromagnet with $J=0$, half-moons have been observed at very low temperature and finite frequency \cite{robert2008}, while pinch points dominate the low-energy physics and are visible in the structure factor (Fig.~\ref{fig:ChK} and Ref.~[\onlinecite{zhitomirsky2008}]). Away from $J=0$, one could have expected the reversed picture: half-moons at $\omega=0$ and pinch points at finite frequency. However, LL dynamics do not show pinch points at any frequency. This is because the LL dynamics is simulated at relatively higher temperatures, where
the flat band couples with the dispersive ones and the pinch points are washed out. 
A clear separation of energies might require a much lower temperature, which is accessible for the Heisenberg antiferromagnet, but not at intermediate and large values of $J$ where simulations, and thus LL dynamics, either order, or are very hard to thermalize.

The same trend is also seen for a pyrochlore lattice, as shown in Fig. \ref{fig:llg_pyrochlore}.
The half-moon and the star patterns are clearly visible in each region. 
The results of both lattices show that the shadow of pinch points can be observed through the excitations in finite-frequency range, i.e., the proximity to Coulomb phase
can be captured through the inelastic neutron scattering experiment.

\section{Real-space picture:\\ magnetic clustering}
 \label{sec:ising}
 
In this section, we will address the real-space picture, accompanying the characteristic patterns in the structure factors.
We will show that half-moons and stars reflect the formation of magnetic clusters. These magnetic clusters can be associated with the cluster of topological charges obtained in the Ising systems, through the analogy of conserved spin introduced in Eq.~(\ref{eq:Mp}), with the topological charge defined in the Ising system~\cite{udagawa2016, rau2016, mizoguchi2017}.
   
\subsection{Comparison with Ising systems}
\label{sec:Ising}

The half-moons and stars in $\mathcal{S}({\bm{q}})$ are also seen in the corresponding Ising model~\cite{udagawa2016, rau2016, mizoguchi2017}: 
\begin{align}
\mathcal{H} & = \sum_{\langle i,j\rangle_{\mathrm{NN} }} \sigma_i^z \sigma_j^z 
+J \sum_{\langle i,j\rangle_{\mathrm{2nd} }} \sigma_i^z \sigma_j^z 
+J \sum_{\langle i,j\rangle_{\mathrm{3rd,a} }} \sigma_i^z \sigma_j^z.  \label{eq:hamiltonain}
\end{align}
For Ising degrees of freedom $\sigma_{i}^{z}$, the Hamiltonian can be re-written in terms of local topological charges on each triangle (for kagome) or tetrahedron (for pyrochlore),
\begin{align}
Q_n = \zeta_n  \sum_{i \in n} \sigma_i^z,
\label{eq:Qp}
\end{align}
with $\zeta_n = +(-)1$ for an upward (downward) triangle/tetrahedron. 
The possible values of charges are $Q_n = \{\pm 3 , \pm 1\}$ for a kagome lattice and $Q_n = \{\pm 4 , \pm 2, 0\}$ for a pyrochlore lattice. 
The Hamiltonian then becomes~\cite{ishizuka2013, udagawa2016, rau2016, mizoguchi2017}
\begin{equation}
\mathcal{H} = \left( \frac{1}{2} - J \right) \sum_n Q_n^2 - J \sum_{\langle n,m \rangle} Q_n Q_m + \mathrm{(const.)}.
\label{eq:hami_charge}
\end{equation}
The vector field defined in Eq.~(\ref{eq:Mp}) and the Hamiltonian form of Eq.~(\ref{Ham_conservedspin}) were natural extensions of these discrete topological charges to continuous degrees of freedom. 
We see in Eq.~(\ref{eq:hami_charge}) that $J$ couples the NN charges. 
$J >0 $ means that same-sign charges attract each other. 

\begin{figure}[t]
\begin{center}
\hspace{0cm}\includegraphics[bb = 0 0 407 508, width=9cm]{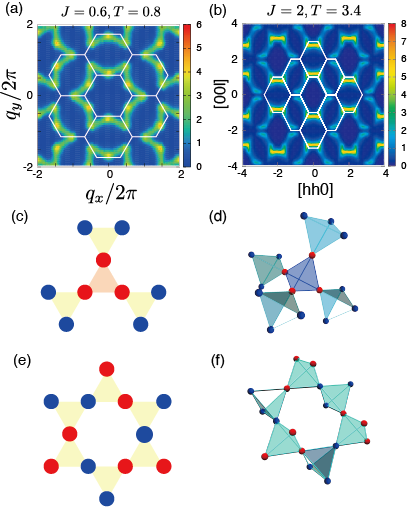}\\
\vspace{-10pt}
\caption{The star patterns in the structure factor for Ising models on the (a) kagome and (b) pyrochlore lattices, obtained by Monte Carlo simulations. 
The parameters $J$ and $T$ are described in the figure. White lines denote the Brillouin zones. 
The corresponding clusters of topological charges are shown in (c) and (d). 
The hexamers for (e) kagome and (f) pyrochlore lattices are also presented.
Red (blue) dots denote the spin up (down). 
Orange (yellow) triangles for a kagome lattice denote $Q = + 3 (+1)$;
dark blue (light blue) tetrahedra for a pyrochlore lattice denote $Q = + 4 (+2)$. 
}
\label{fig:ising_star}
\end{center}
\end{figure}

The static structure factors of this Ising model show similar features to those of the Heisenberg model.
The half-moons appear for $0 < J < \frac{1}{3}$ on kagome \cite{mizoguchi2017}, and for $J \sim \frac{1}{4}$ on pyrochlore \cite{udagawa2016,rau2016}. In both lattices, 
the origin of the half-moons is to a large extent due to the formation of ``hexamers" [Figs. \ref{fig:ising_star}(e) and \ref{fig:ising_star}(f)], which correspond to closed loops made of at least six charges of the same sign. 
Branches of same-sign charges are then attached to these central hexamers, forming disordered spin-liquid phases made of large clusters of topological charges.

Here, we show that the star patterns discussed in this paper also appear in the large-$J$ region of the Ising models: $J > \frac{1}{3}$ for a kagome lattice [Fig.~\ref{fig:ising_star}(a)] and $J > \frac{1}{4}$ for a pyrochlore lattice [Fig. \ref{fig:ising_star}(b)]. At lower temperatures, the system orders into phases tiled by small clusters of charges, with a maximal charge at the center, surrounded by smaller charges of the same sign [Figs. \ref{fig:ising_star}(c) and \ref{fig:ising_star}(d)]. 

From this point of view, the passage from the half-moons to the stars in the static structure factor corresponds to the evolution from a disordered phase made of hexamers to {the ordered phase of} smaller clusters centered around a maximal charge. 
The similarity of half-moons/stars between Ising and Heisenberg models suggests that short-range correlations similar to topological charge clusters also develop in the Heisenberg models, even though topological stability, with a discretized value of the topological charge, cannot be expected for the continuous spin systems.
The motivation of the next sections will be to make this idea more quantitative.

\subsection{Conserved-spin correlator}
\begin{figure}[t]
\begin{center}
\includegraphics[bb = 0 0 422 611, width=\linewidth]{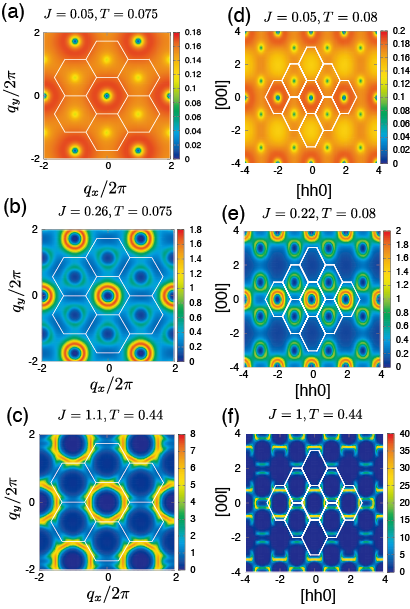}\\
\vspace{-10pt}
\caption{
The conserved spin correlator for a kagome lattice [(a)-(c)] and for a pyrochlore lattice [(d)-(f)]. 
Corresponding values of $J$ and $T$ are given in the figure. 
}
\label{fig:ccq_pyrochlore}
\end{center}
\end{figure}

To characterize
the real-space structure in the Heisenberg models, we focus on
the conserved spin $\bm{M}_n$ as a vector-field analog of the topological charge $Q_n$ in the Ising models. 
The momentum-space correlator of $\bm{M}_n$ is defined as
\begin{align}
S_{\bm{M}}(\bm{q}) \equiv \frac{N_{\mathrm{sub}}}{N_{\mathrm{site}}} \sum_{n,m} \langle \bm{M}_n \cdot \bm{M}_m \rangle e^{-i \bm{q} \cdot (\bm{R}_m -\bm{R}_n)},
\end{align}
where $\bm{R}_{n,m}$ is the coordinate at the center of the triangle/tetrahedron where $\bm{M}_{n,m}$ is defined. 
Within the large-N approximation, $S_{\bm{M}}(\bm{q})$ is represented as
\begin{align}
S_{\bm{M}}(\bm{q}) = & 
\frac{N_{\mathrm{sub}}}{N_{\mathrm{site}}}
\sum_{k, k^\prime = \bigtriangleup, \bigtriangledown} \zeta_{k} \zeta_{k^\prime} \sum_{\eta = 1}^{N_{\mathrm{sub}}} 
\langle M^{\eta}_{k} (-\bm{q}) M^{\eta}_{k^\prime} (\bm{q}) \rangle \notag \\
=  & \sum_{\eta = 1}^{N_{\mathrm{sub}} } \sum_{\mu, \nu} \frac{ [\bm{\psi}^\ast_\eta (\bm{q})]_\mu [\bm{\psi}_\eta (\bm{q})]_\nu }{  [\lambda + \beta \varepsilon_\eta(\bm{q}) ] } \cdot F_{\mu ,\nu}(\bm{q}).
\end{align}
Here, 
$k= \bigtriangleup, \bigtriangledown$ denotes the direction of  triangle/tetrahedron,
$\bm{r}_{c} \equiv \frac{1}{N_{\rm sub}} \sum_{\mu} \bm{r}_\mu $ is the coordinate at the center of the upper triangle/tetrahedron,
\begin{equation}
M^{\eta}_{k} (\bm{q}) = \sum_{\mu} e^{i \bm{q} \cdot [ \zeta_{k} (\bm{r}_\mu- \bm{r}_c)]} [\psi_\eta(\bm{q})]_{\mu},
\end{equation}
is the conserved spin of $\eta$-band, and 
\begin{equation}
F_{\mu ,\nu}(\bm{q}) =  \sum_{k, k^\prime = \bigtriangleup, \bigtriangledown } \zeta_{k} \zeta_{k^\prime} e^{i \bm{q} \cdot [ \zeta_{k^\prime} (\bm{r}_\nu- \bm{r}_c) - \zeta_k  (\bm{r}_\mu- \bm{r}_c)]},   
\end{equation}
is the additional form factor.
Figure \ref{fig:ccq_pyrochlore} shows $S_{\bm{M}}(\bm{q})$
obtained by the large-$N$ approximation. 
In the region (I), $S_{\bm{M}}(\bm{q})$ becomes very small with decreasing temperature,  
due to the divergence-free nature of the Coulomb phase. 
In the large-$N$ sense, the flat band does not contribute to the conserved spin correlator
since $M_{k}^{\mu} (\bm{q}) = 0$ for the flat bands.
For regions (II) and (III), since $S_{\bm{M}}(\bm{q})$ is written by a linear combination of sublattice-resolved structure factors,
it shows the characteristic patterns reminiscent of the static structure factor $\mathcal{S}({\bm{q}})$.
While the shadow pinch points are absent
due to the additional form factor $F_{\mu ,\nu}^{k,k^\prime}(\bm{q})$,
the high-intensity points of $S_{\bm{M}}(\bm{q})$ trace the trajectory of $\varepsilon_{\rm min}(\bm{q})$ in Fourier space. 

\begin{figure}[t]
\begin{center}
\includegraphics[bb = 0 0 485 200,width=\linewidth]{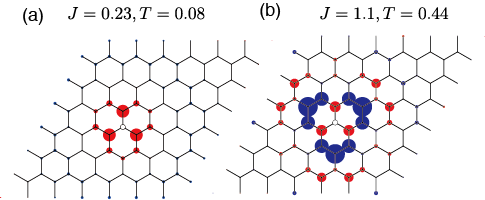}\\
\vspace{-10pt}
\caption{The conserved spin correlator in real space for a kagome lattice.
 Black lines denote bonds of the premedial honeycomb lattice. 
 Colors of dots correspond to the sign (see the main text),
 and the central white circle denotes the origin. 
 The radius is proportional to the absolute value, and rescaled as 
 $r = C |S_{\bm{M}}(\bm{R}) |$ with $C=3$ for (a) and $C =1 $ for (b).
}
\label{fig:ccr_kagome}
\end{center}
\end{figure}
\begin{figure}[t]
\begin{center}
\includegraphics[bb = 0 0 535 215, width=\linewidth]{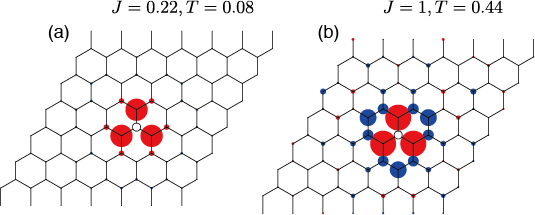}\\
\vspace{-10pt}
\caption{The conserved spin correlator in real space for a pyrochlore lattice. Black lines denote bonds of one buckled honeycomb layer, cut through the premedial diamond lattice, as seen from the [111] direction. 
 Colors of dots correspond to the sign (see the main text),
 and the central white circle denotes the origin.
 The radius is proportional to the absolute value, and rescaled as 
 $r = C |S_{\bm{M}}(\bm{R}) |$ with $C=3$ for (a) and $C =1/3 $ for (b). 
}
\label{fig:ccr_pyrochlore}
\end{center}
\end{figure}

The real-space correlator,
\begin{equation}
S_{\bm{M}}(\bm{R}) \equiv \langle \bm{M}_0 \cdot \bm{M}_{\bm{R}}  \rangle,
\label{eq:rsc} 
\end{equation} 
is defined on dual honeycomb (diamond) lattice for the kagome (pyrochlore) case, and given by the inverse Fourier transformation of $S_{\bm{M}}(\bm{q})$.
We show $S_{\bm{M}}(\bm{R})$ for both cases,
in Figs. \ref{fig:ccr_kagome} (kagome) and \ref{fig:ccr_pyrochlore} (pyrochlore). 
The site $0$ is shown with a white circle.
The red (blue) dot means that the correlation $\langle \bm{M}_0 \cdot \bm{M}_{\bm{R}}  \rangle$ takes a positive (negative) value,
and its radius denotes the rescaled absolute value (see the captions of the figures). 
Since we are interested in the cooperative, but nonetheless disordered, paramagnetic phase where $\langle \mathbf{M}\rangle=0$, the averaged real-space correlator of Eq.~(\ref{eq:rsc}) offers an alternative to the spin-configuration snapshot available in the ordered phase of the Ising model. With continuous spins, the real-space representation of this clustering is further complicated by the continuous evolution of the half-moon radius in the phase diagram, which implies an incommensurate wave-vector ordering for generic values of $J$. This issue can, however, be resolved at high-symmetry points of the Hamiltonian, such as exemplified in Sec. \ref{sec:highsym}.

As shown in Fig.~\ref{fig:ccr_kagome} (a), in the region (II), positive correlations develop in NN sites, 
as can be expected 
from the $\bm{M}_n$-representation of Hamiltonian [Eq. (\ref{Ham_conservedspin})]. 
Moreover, a noticeable correlation develops beyond n.n sites, especially in surrounding hexagons, implying the clustering of spins reminiscent of the hexamer cluster in the Ising case, made of same-sign charges surrounding a hexagon.
In the Heisenberg case, instead of the charge, the conserved spin shows substantial positive correlation around a hexagon.

The qualitative difference of patterns between region (II) with half-moons and region (III) with stars indicates that different types of clusters evolve in these two regions. 
For region (III) the positive NN correlation is surrounded by the negative correlations, which is reminiscent of the crystalization of double- and triple-charge clusters in the Ising models~\cite{udagawa2016,mizoguchi2017}.

Despite the qualitative similarity of cluster structures between Heisenberg and Ising cases, there is one significant difference.
In the Ising case, the cluster shapes are rigidly fixed in regions (II) and (III), respectively, due to the discrete spin nature 
of Ising spins, and do not change with $J$. 
Meanwhile in the Heisenberg case, the continuous spins 
allow continuous modification of clusters, and their amplitudes and cluster sizes also change continuously with $J$. For example, 
in region (II), the cluster can be considerably long-ranged near the boundary with region (I), as implied by the small half-moon radius in Fourier space [Fig.~\ref{fig:radius}].

\subsection{Gauge-charge ordering at the high-symmetry point $J=J_{2c}$ on pyrochlore \label{sec:highsym}}

We next confirm the change of clustering patterns between regions (II) and (III)
via Monte Carlo simulations, using the pyrochlore model at the boundary $J=J_{2c}$ as a working example.

What happens at $J_{2c}$ ? Within the large-$N$ approximation, this is where the energy-minima manifold changes its topology, as illustrated in Figs.~\ref{fig:minima_kagome} and \ref{fig:minima_pyrochlore}. The manifold moves from enclosing the $\Gamma$ point in region (II)  ($J_{1c}<J<J_{2c}$) to the zone corners in region (III) ($J_{2c}<J$). The model at $J_{2c}$ is thus a high-symmetry point of our Hamiltonian. As we will see in this section, it confers to the $J_{2c}$ boundary an advantage of simplicity particularly useful to characterize the low-temperature ordered state.

\begin{figure}[t]
\centering\includegraphics[width=\columnwidth]{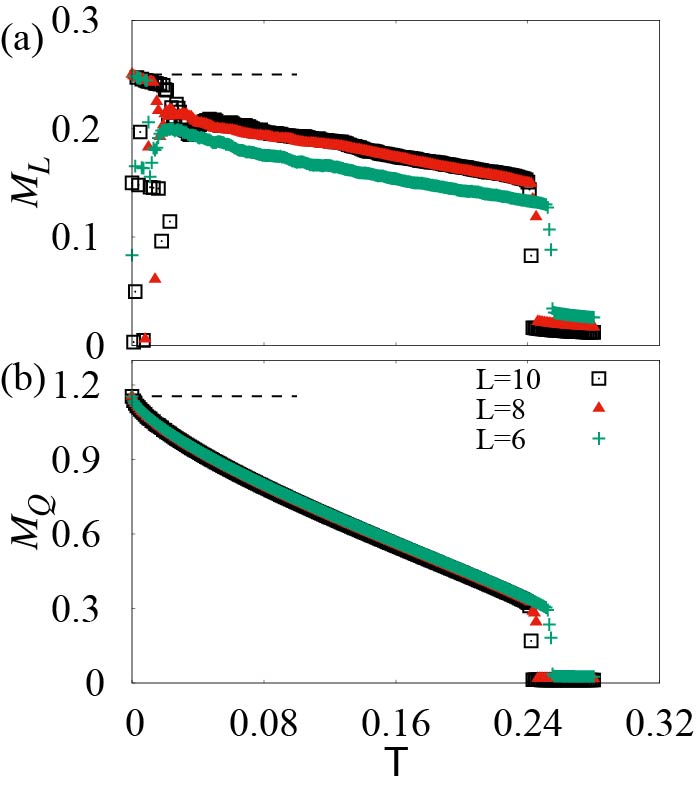}
\caption{
First order phase transition at the high-symmetry point $J=J_{2c}$ on pyrochlore, as demonstrated by the discontinuity of (a) the dipolar order parameters $M_{L}$ at wave vector $\mathbf{q}_{L}$ and (b) the quadrupolar order parameter $M_Q$. The black dashed lines indicate the value of saturation for each order parameter.
}
\label{fig:J05obs}
\end{figure}

Since the energy minima cover an extended region in Fourier space, the ordering mechanism is necessarily via thermal order by disorder. For most of the phase diagram when $J>J_{1c}$, the continuous evolution of the energy-minima manifold implies incommensurate order at low temperatures. But, at the high-symmetry model $J_{2c}$, order by disorder selects the $L$ point on the boundary of the Brillouin zone, (h,k,l)$=(1/2,1/2,1/2)$, as suggested by the large-$N$ analysis of Fig.~\ref{fig:band_py} and confirmed by Bragg peaks in the Monte Carlo of Fig.~\ref{fig:HMP} (bottom right panel). The corresponding order parameter 
\begin{eqnarray}
M_{L}=\left| \dfrac{1}{N}\sum_{i=1}^{N}\mathbf{S}_{i} \textrm{e}^{\imath \mathbf{r}_{i}\cdot \mathbf{q}_{L}}\right|
\label{eq:ML}
\end{eqnarray}
displays a clear first-order jump at the transition (Fig.~\ref{fig:J05obs}). Since there are eight $L$ points for each Brillouin zone with each $L$ point shared between two adjacent Brillouin zones, 
the saturated value of $M_{L}$ is $2/8=1/4$. Measurements of $M_{L}$ are especially difficult to thermalize below the transition. $M_{L}$ seems to converge towards its saturated value of $1/4$, possibly via a second transition at very low temperature. However, further work is necessary to confirm this point. It is also possible that the order parameter $M_{L}$ does not saturate. Measurements of the quadrupolar order parameter $M_{Q}$, on the other hand, thermalize quite easily to its saturated value of $2/\sqrt{3}$ at zero temperature (see Appendix~\ref{app:Q} for the definition).

\begin{figure*}[t]
\centering\includegraphics[width=18cm]{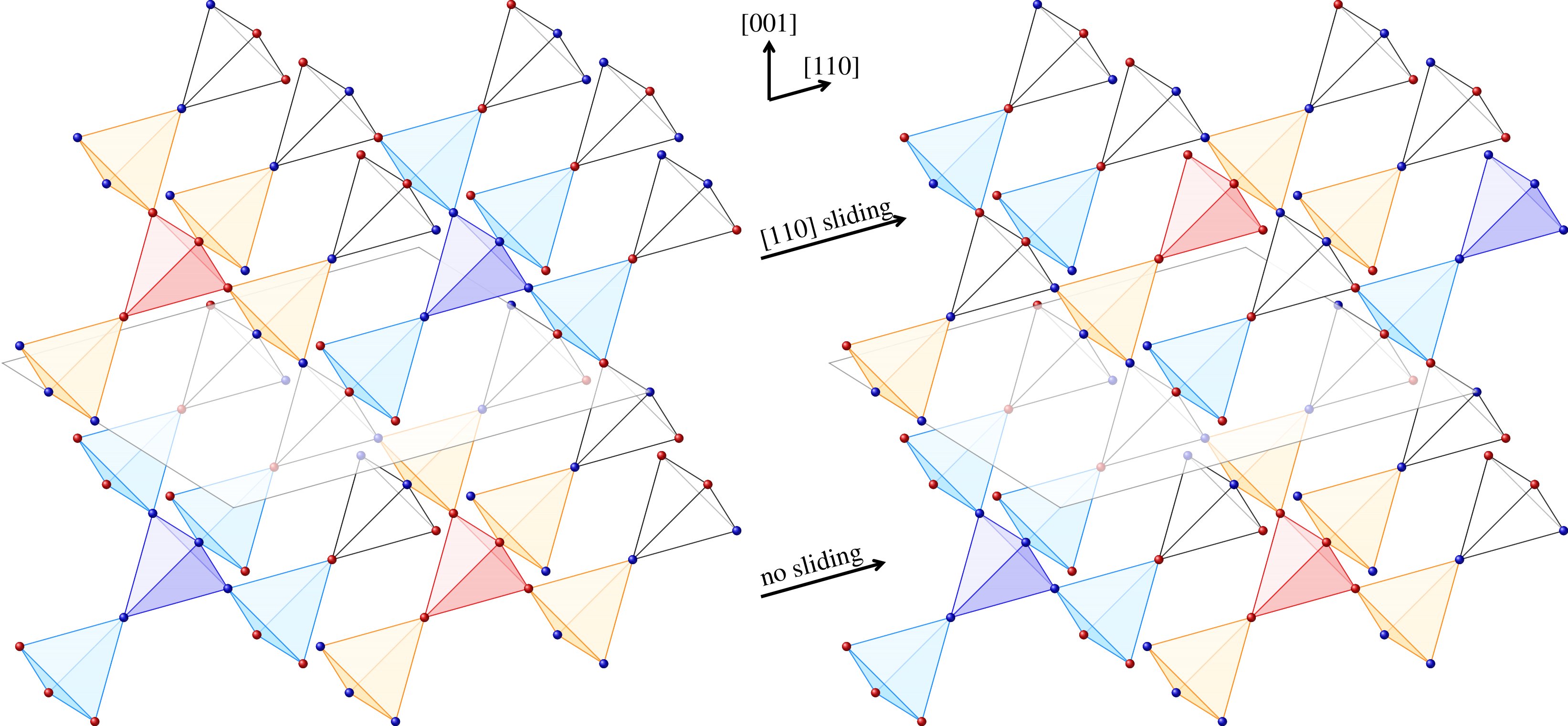}
\caption{Ground state at the boundary $J=J_{2c}$ on the pyrochlore lattice displayed over 32 tetrahedra. All spins are collinear, pointing either up (blue sphere) or down (red sphere). The color of the tetrahedra represent their effective topological charges as defined in Eq.~(\ref{eq:Qp}): $+4$ (dark blue), $+2$ (light blue), $0$ (white), $-2$ (orange), $-4$ (red). Each cubic unit cell is composed of a small cluster of five charges (including one maximal charge) and three zero charges (acting as a vacuum separating charges of opposite sign). This intervening vacuum allows for a global shift of the clusters, illustrated between the left and right panels. The spin configurations are split into two parts by an horizontal semi-transparent plane. Between the two panels, all spins above the semi-transparent plane have been shifted along the [110] direction to the next tetrahedron (at a distance of two nearest neighbors). Because the contact between charges is not modified, the energy is the same. Doing the same shift a second time gives back the initial state. From the point of view of discrete degrees of freedom, this gives a sub-extensive entropy to the ground state.
}
\label{fig:meth}
\end{figure*}

In the rest of this section, we will focus on the nature of the ground state. The saturation of the quadrupolar order parameter $M_{Q}$ [Fig.~\ref{fig:J05obs}.(b)] indicates that all spins are collinear in the ground state. This collinearity allows to temporarily forget the continuous nature of the classical Heisenberg spins and to consider them as Ising degrees of freedom, pointing either up or down. In analogy with Sec.~\ref{sec:Ising}, each tetrahedron bears an effective charge $Q_{n}=\{0,\pm 2,\pm 4\}$. At $J_{2c}=1/2$, the ``chemical potential" (i.e. the first term) of Eq.~(\ref{eq:hami_charge}) disappears and the ground-state energy, $E_{GS}$, only comes from the NN coupling between charges
\begin{eqnarray}
E_{GS} = - \dfrac{1}{2} \sum_{\langle n,m \rangle} Q_n Q_m .
\label{eq:EGSJ05}
\end{eqnarray}
This coupling is attractive between same-sign charges. Because of the staggering of $\zeta_{n}$ between up and down tetrahedra in Eq.~(\ref{eq:Qp}), this attraction does not give rise to ferromagnetism. The minimization of the energy is done by tiling the lattice with the small clusters of Fig.~\ref{fig:ising_star}(d): a central charge $+ 4$ or $-4$ surrounded by four charges $+2$ or $-2$, respectively. This state is illustrated in Fig.~\ref{fig:meth} over four cubic unit cells (32 tetrahedra). In order to avoid contact between charges of opposite sign, each cluster is separated from the other ones by a vacuum (zero-charge tetrahedra). 
Furthermore, each cluster fits within the eight tetrahedra of a cubic unit cell: one maximal charge $Q_{n}=\pm 4$, four charges $Q_{n}=\pm 2$ and three vacuum tetrahedra $Q_{n}=0$. In order to respect the global neutrality of the system, there must be as many positive as negative clusters. Since the centers of the cubic unit cells form, by definition, a bipartite cubic lattice, the global neutrality is enforced by a staggering arrangement of the clusters, alternatingly positive and negative. The magnetic unit cell of the ground state is made of 32 spins (16 tetrahedra).

This structure is responsible for the peaks at (h,k,l)$=(\frac{1}{2},\frac{1}{2},\frac{1}{2})$ in the structure factor of Fig.~\ref{fig:HMP} (bottom right panel). 
In the actual simulations, however, 
the difficulty of thermalization remains at low temperatures,
because the order parameter $M_{L}$ does not saturate completely. 
We believe this is due to the sub-extensive entropy of the ground state, as illustrated in Fig.~\ref{fig:meth}. We explain the origin of the degeneracy of the $(frac{1}{2},\frac{1}{2},\frac{1}{2})$-state in Appendix \ref{app:order}. 

\section{Summary and discussion} 
\label{sec:summary}
We have investigated the magnetic correlations of Heisenberg models with antiferromagnetic farther-neighbor interactions on kagome and pyrochlore lattices in their cooperative paramagnetic regions. For both lattices, we found three distinct patterns of the static structure factor, $\mathcal{S}(\bm{q})$, namely pinch points, half-moons, and stars. We clarified the origin of these patterns by combining the band structure analysis based on a large-$N$ approximation, and Monte Carlo simulations.

Among the above characteristic patterns of the structure factor, the pinch point serves as a direct evidence of a Coulomb phase. Vicinity to a Coulomb phase is signaled by the presence of
half-moon patterns. This  can be interpreted as complementary to pinch points: they live in the dark regions of
the pinch points, and unlike those, generally incorporate a dispersion, so that their radius (distance from the pinch point) changes with energy. As their radius increases, they eventually connect with half moons from neighboring Brillouin
zones to generate the star patterns (Figs.~\ref{fig:minima_kagome} and \ref{fig:minima_pyrochlore}).
Depending on the relative ordering of the bands in energy, these features may even appear in the ground-state correlations.

From a real-space perspective, half-moon and star patterns reflect the formation of magnetic clusters. These clusters involve short-range correlation of the conserved spin, which is analogous to the topological charge defined for the corresponding Ising system. Through this analogy, the half-moon and star cluster can be associated with hexamers and triple-charge clusters obtained in the Ising system, respectively.

The analogy to Ising systems can be extended to the ordering at the high-symmetry point, $J = J_{2c}$ for the pyrochlore system, where the structure of the low-temperature ordered phase can be well understood through the concept of topological charge. In contrast, the difference from the Ising system is found in the rigidity of the cluster structure: While the clusters are rigidly fixed due to the discreteness of the spins in the Ising system, in the Heisenberg system, the cluster shape is flexibly changed upon varying $J$, due to the continuous nature of spin degrees of freedom.

We further addressed the dynamical properties of the model by solving the semiclassical LL equations. As a result, we found that the characteristic half-moon and star patterns appear in the frequency-resolved structure factors, in particular in the low-frequency regime, which means the magnetic clusters dominate the long-time behavior of the dynamics. The patterns in $\mathcal{S}(\bm{q},\omega)$ presented here can be directly accessed through experimental probes, such as inelastic neutron scattering.

In fact, in pyrochlore compounds, several types of magnetic clusterings have been reported. In ZnCr$_2$O$_4$~\cite{lee2002} and MgCr$_2$O$_4$~\cite{tomiyasu2008,tomiyasu2013,gao2018}, six-spin composites dominate the low-energy excitations. While the proposed spatial structure is different from the hexamers obtained in our analysis, our model clearly gives a route to similar clustering around hexagons. It is tempting to point out the possibility that our hexamers may be continuously connected to the low-energy excitations observed for these materials. In this respect, a molybdate pyrochlore material Lu$_2$Mo$_2$O$_5$N$_2$ will also provide an interesting perspective~\cite{clark2014,iqbal2017}. As for the shadow of pinch points in dispersive bands, it has also been observed in a kagome model for Fe jarosites with Dzyaloshinskii-Moriya and second-neighbor interactions~\cite{chernyshev2015}. Potential connections to half-moon and star patterns in this material have not been investigated yet, and would be an interesting direction to follow. On the other hand, the half-moon signal corresponding to the hexamer-type clusters was recently theoretically proposed for a double-layered kagome material~\cite{pohle2017}. 

It is also worthwhile to look at another pyrochlore compound, ZnFe$_2$O$_4$. The cluster excitation observed for this compound takes a ``dodecamer'' form, consisting of 12 spins~\cite{tomiyasu2011_2}. The spatial structure of this excitation is the same as the triple-charge cluster obtained in the region (III) in our model. Interestingly, for this compound, a large farther-neighbor coupling, $J_{3a} > J_1$, is expected~\cite{yamada2002,kamazawa2003,tomiyasu2011_2}. A different dodecamer structure reminiscent of the kagome hexamer was also proposed for HgCr$_2$O$_4$~\cite{tomiyasu2011}.

In conclusion, we found new characteristic patterns in magnetic structure factors, complementary to pinch points, which signal the proximity to a Coulomb phase. 
These patterns signal the formation of magnetic clusters, analogous to the low-energy excitations observed for pyrochlore compounds.

\acknowledgements
We thank Keisuke Tomiyasu and Jeffrey G. Rau for interesting discussions.
This work was supported by the JSPS KAKENHI (Grants No. JP15H05852, No. JP16H04026 and No. JP26400339), MEXT, Japan,
and by the Deutsche Forschungsgemeinschaft under Grant No. SFB 1143. 
TM wishes to thank Max-Planck-Institut fur Physik komplexer Systeme, 
where part of the present work was done.
L.D.C.J. acknowledges support from the University of Bordeaux (IdEx BIS) and from the TQM Unit of the Okinawa Institute of Science and Technology Graduate University, as well as hospitality from Gakushuin University in Tokyo. Part of numerical calculations were carried out on the Supercomputer Center at Institute for Solid State Physics, University of Tokyo.

\textit{Note added:} Recently, we learned of a parallel study by Yan {\it et al}., which reports complementary results for a different model \cite{Yan2018}.

\appendix
\section{Analytical formula for $\mathcal{S}(\bm{q})$ in large-$N$ analysis}
\label{sec:largeNanalytics}
In this appendix, we describe the derivation of $\mathcal{S}(q)$ in large-$N$ approximation, in detail. 
The $J_1$-$J_2$-$J_3$ model has two special properties at $J_2=J_3\equiv J$, which enables us simple analytical approach. 
One is the polynomial expression of Hamiltonian in terms of incident matrices, and the other is the line graph correspondence. 
With the help of graph-theoretical argument, we can construct a simple and systematic way to obtain the analytical expression of $\mathcal{S}(\bm{q})$ in Eq. (\ref{eq:ssf_diag}).

\subsection{Polynomial expression}
\label{sec:polynomial}
We consider kagome and pyrochlore lattices on the same footing, 
and start with introducing an $N\times N$ incident matrix, $\hat{\delta}^{(1)}\equiv\hat{h}$,
where we write $N = N_{\rm site}$ for brevity.
Each row $j$, and column indices $j'$ correspond to the sites of the lattice, and the matrix element takes
\begin{align}
[\hat{\delta}^{(1)}]_{jj'} = [\hat{h}]_{jj'} = \left\{\begin{array}{ll}
1 & {\rm if}\ j\ \&\ j' {\rm are\ connected}\\
0 & {\rm otherwise}
\end{array}\right.
\end{align}
The Hamiltonian matrix can be expressed as $\hat{H}=\hat{h}$ at $J=0$, supposing $J_1=1$ and $J_2=J_3=J$.

Generalizing $\hat{\delta}^{(1)}$, we introduce a matrix, $\hat{\delta}^{(n)}$, so that the element $[\hat{\delta}^{(n)}]_{jj'}$ takes $1$, if and only if the two sites, $j$ and $j'$ are $n$ Manhattan distance away, and otherwise, $0$. 
Note that for the kagome and pyrochlore lattices, the Manhattan distance is the minimal number of NN bonds necessary to connect two sites.

Since the squared incident matrix $(\hat{\delta}^{(1)})^2$ connects any two sites, where one can be reached from the other in two hoppings, one can obtain
\begin{align}
(\hat{\delta}^{(1)})^2 = z\hat{I}_{N \times N} + x \hat{\delta}^{(1)} + \hat{\delta}^{(2)}, \label{eq:2ndhopping}
\end{align}
with $z=2(N_{\rm sub} - 1)$ is a number of coordination,
and $x = N_{\rm sub} - 2$ is a number of paths 
through which, starting from a site, one reaches a NN site of that site
by two other NN moves. 
Hereafter, $\hat{I}_{\ell \times \ell }$ represents the $\ell \times \ell$ identity matrix. 

Obviously, $\hat{\delta}^{(2)}$ corresponds to the part of Hamiltonian matrix describing the second- and third-neighbor interactions, so we can express
 \begin{align}
\hat{H} &= \hat{\delta}^{(1)} + J\hat{\delta}^{(2)} = \hat{\delta}^{(1)} + J(\hat{\delta}^{(1)})^2 
- xJ \hat{\delta}^{(1)}
- zJ\hat{I}_{N \times N}
\nonumber\\
&= (1-xJ) \hat{h} + J\hat{h}^2 - zJ \hat{I}_{N \times N}. \label{eq:poly_1}
\end{align}
Now the Hamiltonian matrix is expressed as a polynomial of incident matrix, $\hat{h}$,
the eigenvalue problem of $\hat{H}$ is reduced to that of $\hat{h}$.

\subsection{Dual lattice}
\label{sec:duallattice}
\begin{figure}[t]
\begin{center}
\hspace{-1cm}\includegraphics[bb = 0 0 342 473, width=8cm]{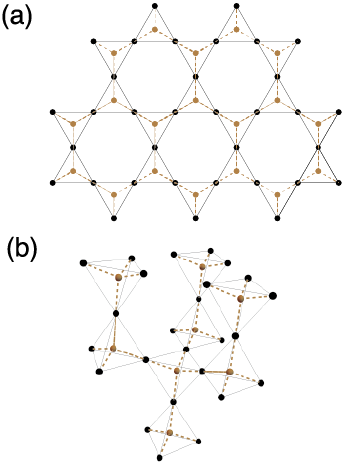}\\
\vspace{-10pt}
\caption{Dual lattices for (a) kagome and (b) pyrochlore lattices.
Black dots/spheres denote original lattices, and brown ones denote dual lattices.}
\label{fig:duallattice}
\end{center}
\end{figure}

To solve the eigenvalue problem of $\hat{h}$, it is convenient to introduce dual lattice. 
For clarity, we focus on a kagome lattice, first.
We start with constructing an intermediate lattice, by placing new sites on the centers of triangles, and connecting the new sites and neighboring old sites, while erasing the original bonds of kagome lattice.
Secondly, from this intermediate lattice, we erase the original sites of the kagome lattice, and obtain a honeycomb lattice as a dual lattice  [Fig. \ref{fig:duallattice}(a)].
As a dual lattice of pyrochlore lattice, we obtain a diamond lattice in a similar way [Fig. \ref{fig:duallattice}(b)]. 

The dual lattice shares the same unit cell as the original lattice.
Below, we adopt the following lattice conventions.
For a kagome lattice, as lattice vectors, we choose $\bm{a}^{(\mathrm{K})}_1 = (1,0)$ and $\bm{a}^{(\mathrm{K})}_2 = \left( \frac{1}{2},\frac{\sqrt{3}}{2} \right)$,
and as the coordinates of three sublattices, 1,2, and 3, $\bm{r}^{(\mathrm{K})}_1 = \left( 0,0 \right)$, $\bm{r}^{(\mathrm{K})}_2 = \left( \frac{1}{4} , \frac{\sqrt{3}}{4} \right)$, $\bm{r}^{(\mathrm{K})}_3 = \left(  \frac{1}{2} , 0 \right)$. 
Accordingly, the coordinates of two sublattices, A and B of the dual honeycomb lattice are $\bm{r}^{(\mathrm{H})}_{\mathrm{A}} = \left( \frac{5}{4}, \frac{5\sqrt{3}}{12}\right)$ and $\bm{r}^{(\mathrm{H})}_{\mathrm{B}} = \left( \frac{1}{4} , \frac{\sqrt{3}}{12} \right)$. 
For a pyrochlore lattice,
the lattice vectors are
$\bm{a}^{\mathrm{P}}_1 = \left(0, \frac{1}{2} , \frac{1}{2} \right)$,
$\bm{a}^{\mathrm{P}}_2 = \left( \frac{1}{2} , 0, \frac{1}{2} \right)$,
$\bm{a}^{\mathrm{P}}_3 = \left( \frac{1}{2} , \frac{1}{2}, 0 \right)$,
and the positions of four sublattices are
$\bm{r}^{\mathrm{P}}_1 = \left(0,0,0 \right)$,
$\bm{r}^{\mathrm{P}}_2 = \left(0,\frac{1}{4},\frac{1}{4} \right)$,
$\bm{r}^{\mathrm{P}}_3 = \left(\frac{1}{4},0,\frac{1}{4}\right)$,
$\bm{r}^{\mathrm{P}}_4 = \left(\frac{1}{4},\frac{1}{4},0 \right)$.
For the dual diamond lattice, the coordinates of two sublattices (A and B) are
$\bm{r}^{\mathrm{D}}_A = \left( \frac{1}{8} , \frac{1}{8} , \frac{1}{8}  \right)$, and 
$\bm{r}^{\mathrm{D}}_B = \left( \frac{7}{8} , \frac{7}{8} , \frac{7}{8}  \right)$.

\subsection{Line-graph correspondence \label{sec:linegraph}}
\label{sec:linegraph}
Here, let us apply the idea of dual lattice to solve the eigenvalue problem of $\hat{h}$. 
Here, we focus on a kagome lattice, again.
First, we look at the intermediate lattice we have introduced in the previous subsection. 
On this graph, we introduce $N\times N_{\rm H}$ rectangular matrix, 
$\hat{h}^{\rm K\leftarrow H}$, whose $N$ rows correspond to sites on a kagome lattice, and $N_{\rm H}$ columns correspond to the sites of a honeycomb lattice.
We define $\hat{h}^{\rm K\leftarrow H}$ as an incident matrix for the intermediate lattice, i.e., we set
\begin{align}
[\hat{h}^{\rm K\leftarrow H}]_{jl} = \left\{\begin{array}{ll}
1 & {\rm if}\ j\ \&\ l\ {\rm are \ connected}\\
0 & {\rm otherwise}
\end{array}\right.
\end{align}
And we define $N_{\rm H}\times N$ rectangular matrix, $\hat{h}^{\rm H\leftarrow K}$ as $\hat{h}^{\rm H\leftarrow K} = (\hat{h}^{\rm K\leftarrow H})^t$.

The key step to solve the eigenvalue problem is the observation that the matrix $\hat{h}$ can be written as a product of $\hat{h}^{\rm H\leftarrow K}$ and $\hat{h}^{\rm K\leftarrow H}$:
\begin{align}
\hat{h} = \hat{h}^{\rm K\leftarrow H}\hat{h}^{\rm H\leftarrow K} - 2\hat{I}_{N \times N}. 
\label{eq:Hamiltonian_decomposition}
\end{align}
The form (\ref{eq:Hamiltonian_decomposition}) immediately tells us significant information on the energy spectrum of $\hat{h}$.
For $N\times M$ matrix $\hat{A}$ and $M\times L$ matrix $\hat{B}$, it is known that
\begin{itemize}
\item[(i)] rank$\hat{A}$ $\leq$ min\{$N$, $M$\} 
\item[(ii)] rank$\hat{A}\hat{B}$ $\leq$ min\{rank$\hat{A}$, rank$\hat{B}$\} 
\end{itemize}
Applying these properties to $\hat{h}^{\rm K\leftarrow H}$ and $\hat{h}^{\rm H\leftarrow K}$, we obtain
\begin{align}
{\rm rank}(\hat{h}^{\rm K\leftarrow H}\hat{h}^{\rm H\leftarrow K})\leq N_{\rm H} = (2/3)N.
\end{align}
This inequality results in the existence of at least $N-N_{\rm H}=\frac{1}{3}N$ zero modes, i.e. $\hat{h}$ has $\frac{1}{3}N$ eigenstates with degenerate eigenenergy, $-2$. Moreover, the matrix $\hat{h}^{\rm K\leftarrow H}\hat{h}^{\rm H\leftarrow K}$ and the inverse product, $\hat{h}^{\rm H\leftarrow K}\hat{h}^{\rm K\leftarrow H}$ share the common non-zero eigenvalues. Accordingly, given that the incident matrix of honeycomb lattice is given by
\begin{align}
\hat{h}^{\rm H} = \hat{h}^{\rm H\leftarrow K}\hat{h}^{\rm K\leftarrow H} - N_{\rm sub} \hat{I}_{N^{\mathrm{H}} \times N^{\mathrm{H}}}, \label{eq:poly4}
\end{align}
the eigenspectrum of $\hat{h}$ consists of $N_{\rm H}$ eigenvalues of $\hat{h}^{\rm H} + (N_{\rm sub} - 2)\hat{1}$, and $N-N_{\rm H}$-fold degenerate modes with eigenvalue, $-2$.

\subsection{Momentum-space expression \label{sec:mom}}
The translational invariance of the Hamiltonian matrix 
allows us to block-diagonalize it with respect to the momentum $\bm{q}$.  
For each $\bm{q}$, we obtain
$N_{\rm sub}\times N_{\rm sub}$ Hamiltonian matrix $\hat{H}(\bm{q})$ defined in Eq. (\ref{eq:hami_matrix}).
Due to the polynomial expression in Eq. (\ref{eq:2ndhopping}),
we obtain
\begin{equation}
\hat{h}_2 (\bm{q}) = [\hat{h}_1(\bm{q})]^2 - x \hat{h}_1(\bm{q}) - z  \hat{I}_{N_{\rm sub}\times N_{\rm sub}}, \label{eq:poly_3}
\end{equation}
with 
\begin{align}
[\hat{h}_i (\bm{q})]_{\mu \nu} = \sum_{m} [\delta^{(i)}]_{(0,\mu),(m,\nu)} e^{-i \bm{q} \cdot (\bm{R}_m + \bm{r}_\nu -\bm{r}_{\mu})}. 
\end{align} 
Therefore, $\hat{H}(\bm{q})$ is also expressed as a polynomial of the Fourier transformation of $\hat{h}_1$:
\begin{align}
\hat{H}(\bm{q}) = (1-xJ) \hat{h}_1(\bm{q}) + J [\hat{h}_1(\bm{q})]^2 - z J \hat{I}_{N_{\rm sub}\times N_{\rm sub}}, \label{eq:poly_2}
\end{align}
From Eq. (\ref{eq:poly_2}), we see the eigenvalue problem for $\hat{H}(\bm{q})$ is reduced to that for $\hat{h}_1(\bm{q})$.
To solve it, one can utilized the momentum space version of Eq. (\ref{eq:Hamiltonian_decomposition}), namely 
\begin{align}
\hat{h}_1 (\bm{q}) = \hat{h}^{\rm K\leftarrow H} (\bm{q}) \hat{h}^{\rm H\leftarrow K}  (\bm{q}) - 2\hat{I}_{N_{\rm sub}\times N_{\rm sub}}. \label{eq:hami_decomp_2}
\end{align}
Here $\hat{h}^{\rm K\leftarrow H} (\bm{q}) $ is $ N^{\rm H}_{\rm sub} \times N_{\rm sub}$ matrix:  
\begin{align}
[\hat{h}^{\rm K\leftarrow H} (\bm{q})]_{\mu,\nu} = \sum_{m}^{N^{\mathrm{H} }_{\rm u.c.}} 
[\hat{h}^{\rm K\leftarrow H} ]_{(0,\mu) (m,\nu)} e^{-i \bm{q} \cdot (\bm{R}_{m} + \bm{r}_\nu^{\mathrm{H} } - \bm{r}_{\mu})}, 
\end{align} 
and 
\begin{align}
\hat{h}^{\rm H \leftarrow K} (\bm{q}) = [\hat{h}^{\rm K \leftarrow H}(\bm{q}) ] ^{\dagger}. 
\end{align}
Using Eq. (\ref{eq:hami_decomp_2}) as well as the momentum-space analog of Eq. (\ref{eq:poly4}),
one can show that the eigenvalues of $\hat{h}_1 (\bm{q})$ 
consists of eigenvalues of $\hat{h}^{\rm H} (\bm{q})$ (up to a constant) 
and $(N_{\rm sub} - N_{\rm sub}^{\mathrm{H}})$ flat mode with the eigenvalue $-2$; 
$\hat{h}^{\rm H} (\bm{q})$ is given by 
\begin{align}
[\hat{h}^{\rm H} (\bm{q})]_{\mu,\nu} = \sum_{m}^{N_{\rm u.c.}^{\rm H}}  [\hat{h}^{\rm H}]_{(0,\mu),(m,\nu)} e^{-i \bm{q} \cdot (\bm{R}_m + \bm{r}^{\rm H}_\nu -\bm{r}^{\rm H}_{\mu})}. 
\end{align}
Note that the similar relations hold in the case of a pyrochlore lattice. 

\subsection{Kagome lattice \label{sec:appendix_kagome}}
Utilizing the above idea, 
we now show the explicit forms of the eigenvalues and eigenvectors of our model.
Let us first consider a kagome lattice. 
As discussed in the previous section,
the exchange matrix on a kagome lattice is 
expressed as
\begin{equation}
\hat{H}^{\mathrm{K}}(\bm{q}) = \hat{h}^{\mathrm{K}}_1 (\bm{q}) + J \hat{h}^{\mathrm{K}}_2 (\bm{q}), \label{eq:exK}
\end{equation}
with 
\begin{subequations}
\begin{eqnarray}
[\hat{h}^{\mathrm{K}}_1 (\bm{q})]_{\mu \nu} =\left\{ \begin{array}{ll}
2 \cos\bm{q}\cdot (\bm{r}^{\mathrm{K}}_\mu-\bm{r}^{\mathrm{K}}_\nu)&(\mu \neq \nu)\\
0&(\mu= \nu)\\
\end{array} \right.
\end{eqnarray}
\begin{eqnarray}
[\hat{h}^{\mathrm{K}}_2 (\bm{q})]_{\mu \nu} = \left\{ \begin{array}{ll}
2 \cos \bm{q} \cdot \left(\sum_{\rho \neq \mu, \nu}  \bm{r}^{\mathrm{K}}_\mu + \bm{r}^{\mathrm{K}}_\nu -2\bm{r}_\rho^{\mathrm{K} } \right) & (\mu \neq \nu) \\
2 \cos \bm{q} \cdot \left[ 2 \sum_{\rho \neq \mu }  (\bm{r}^{\mathrm{K}}_\rho - \bm{r}^{\mathrm{K}}_\mu) \right] &(\mu = \nu) \\
\end{array} \right. 
\nonumber \\
\end{eqnarray}
\end{subequations}

To obtain the eigenvalues of $\hat{h}^{\mathrm{H}}_{\mathrm{1} }(\bm{q} )$, we first write down the exchange matrix on a dual honeycomb lattice:
\begin{equation}
\hat{h}^{\mathrm{H}}_{\mathrm{1} }(\bm{q} )
= \left(
\begin{array}{cc}
0& G(\bm{q})\\
G^{\ast}(\bm{q}) &0\\ 
\end{array}
\right), 
\end{equation}
with 
\begin{equation}
G(\bm{q}) =  e^{i\frac{q_y}{\sqrt{3}}} + 2 e^{-i\frac{q_y}{2 \sqrt{3}}} \cos \frac{q_x}{2}, 
\end{equation}
The eigenvalues of $\hat{h}^{\mathrm{H}}_{\mathrm{1} }(\bm{q} )$
are given by $\varepsilon_{\pm}^{\rm(H)} (\bm{q}) = \pm |G(\bm{q})|$
and the corresponding eigenvectors are 
\begin{equation}
\bm{\psi}^{\mathrm{H} (\pm)} (\bm{q} ) = \frac{1}{\sqrt{2}} 
\left(
\begin{array}{c}
\pm e^{i\theta_G(\bm{q})/2}  \\
e^{-i\theta_G(\bm{q})/2} \\
\end{array}
\right),
\end{equation}
with $\theta_G(\bm{q}) = \mathrm{arg} G(\bm{q})$.

Then, we immediately obtain the eigenvalues and eigenvectors of (\ref{eq:exK}) 
in the following manner.
First, the eigenvalues of $h^{\mathrm{K}}_{\mathrm{1} }(\bm{q} )$ are identical with those of $h^{\mathrm{H}}_{\mathrm{1} }(\bm{q})$ up to a constant,
and thus,
using Eq. (\ref{eq:poly_3}) and the fact that $z=4, x=1$ for a kagome lattice, 
the two eigenvalues of (\ref{eq:exK}) are obtained as 
\begin{equation}
\varepsilon^{\mathrm{K}}_\pm(\bm{q}) = J |G(\bm{q})|^2  \pm (1+J) |G(\bm{q})| + 1-4J. \label{eq:kagome_eigenvalue}
\end{equation}
Next, to obtain the corresponding eigenvectors, 
we consider a rectangular matrix
$\hat{h}^{\mathrm{K} \rightarrow \mathrm{H}  } (\bm{q})$
as we have discussed int Sec. \ref{sec:linegraph}. 
Its explicit form is given as
\begin{equation}
\hat{h}^{\mathrm{K} \rightarrow \mathrm{H}  } (\bm{q}) =
\left(
\begin{array}{cc} 
e^{i \varphi_1} & e^{-i \varphi_1}\\
e^{i \varphi_2} & e^{-i \varphi_2}\\
e^{i \varphi_3} & e^{-i \varphi_3}\\
\end{array}
\right), 
 \end{equation}
with $\varphi_1 = \frac{q_x}{4} + \frac{q_y}{4\sqrt{3}}$, $\varphi_2 =- \frac{q_y}{2\sqrt{3}}$,and  $\varphi_3 = - \frac{q_x}{4} + \frac{q_y}{4\sqrt{3}}$.
Then, the eigenvectors are obtained as 
\begin{equation}
\bm{\psi}^{\mathrm{K}}_\pm (\bm{q}) = \frac{ h^{\mathrm{K} \rightarrow \mathrm{H}}(\bm{q} )  \bm{\psi}^{\mathrm{H}}_\pm (\bm{q}) }{||  h^{\mathrm{K} \rightarrow \mathrm{H}}(\bm{q} )  \bm{\psi}^{\mathrm{H} }_\pm(\bm{q}) || }.
\end{equation}
Note that the remaining flat mode is orthogonal to two dispersive modes,
and its eigenvalue is $-2(1-J)$.

{\it Energy minima.-}  
We now obtain the eigenvalues of (\ref{eq:exK}), so let us discuss the properties of obtained band structure.
The energy minima of the lower dispersive band can be found by solving 
\begin{equation}
\frac{\partial \varepsilon^{\mathrm{K}}_-(\bm{q})}{\partial q_x} = \frac{\partial \varepsilon^{\mathrm{K}}_-(\bm{q})}{\partial q_y} = 0,  \label{eq:minimacond}
\end{equation} 
By using the expression (\ref{eq:kagome_eigenvalue}), 
one obtains
\begin{equation}
\frac{\partial \varepsilon^{\mathrm{K}}_-(\bm{q})}{\partial q_i} = \left[ 2J |G(\bm{q})| -(1+J) \right]  \frac{\partial |G(\bm{q})|  }{\partial q_i}.
\end{equation} 
Therefore, (\ref{eq:minimacond}) is satisfied when
\begin{enumerate}
\item[(i)] $|G(\bm{q})| = \frac{1+J}{2J} $,
\item[(ii)] $\frac{\partial |G(\bm{q})|}{\partial q_x}  = \frac{\partial |G(\bm{q})|}{\partial q_y} =0 $.
\end{enumerate}
Condition (ii) is satisfied at $\Gamma$, $K$, and $M$ points, but they do not become energy minima.
So let us examine (i).
Notice that the solution of (i)
in $\bm{q}$ space forms lines, rather than a set of discrete points. 
The solution evolves as follows.
First, when $0\leq J\leq \frac{1}{5}$, (i) does not have solutions:
in this region the flat band has the lowest energy and the static structure factors is determined by a flat band,
which gives rise to pinch points in $\mathcal{S}(\bm{q})$.
Second, when $\frac{1}{5} \leq J \leq 1$, 
the solution is given by a closed path enclosing $\Gamma$ point [see Fig. \ref{fig:minima_kagome}(a)]. 
Finally, when $J \geq 1$, 
the solution is given by a closed path enclosing $K$ points [see Fig. \ref{fig:minima_kagome}(b)]. 
As we have seen in the main text, the shape of energy minima 
is reflected to the characteristic shape of the static structure factor, namely, half-moons and stars.

{\it{Derivation of phase boundaries.- }}
\begin{figure}[t]
\begin{center}
\hspace{-1cm}\includegraphics[bb = 0 0 501 250,width=\linewidth]{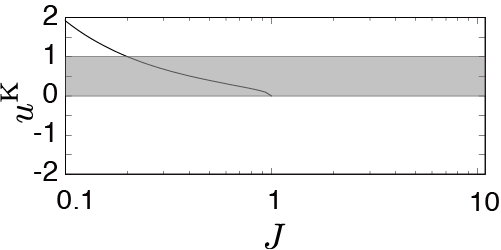}\\
\vspace{-10pt}
\caption{ $J$ dependence of $u^{\mathrm{K}}$.
The solution of Eq. (\ref{eq:min_sol_K}) for $0 \leq Q \leq \pi$ exists for 
the shaded area. 
Note that $u^{\mathrm{K}}$ becomes a complex number for $J \geq 1$. }
\label{fig:boundary_K}
\end{center}
\end{figure}
The change of topology of energy minima surface can be detected
by looking at $\Gamma$M line. 
On $\Gamma$M line, 
we can parametrize the momentum as $\bm{q} = (Q, \frac{Q}{\sqrt{3}}) $ with $0\leq Q \leq \pi$.
Then, $|G(\bm{q})|$ is given by 
\begin{equation}
|G(\bm{q})| = \sqrt{8\cos^2 \frac{Q}{2} +1}, \label{eq:min_sol_K}
\end{equation}
and the solution of (i) is then given by
\begin{equation}
Q = 2 \cos^{-1} \sqrt{\frac{1}{8} \left[ \left( \frac{1+J}{2J} \right)^2-1 \right]}. \label{eq:sol_K}
\end{equation}
Figure \ref{fig:boundary_K} shows $u^{\mathrm{K}} \equiv \sqrt{\frac{1}{8} \left[ \left( \frac{1+J}{2J} \right)^2-1 \right]}$
as a function of $J$. 
In order that $Q$ is between 0 and $\pi$, $u^{\mathrm{K}}$ has to be between $0$ and 1, 
which is represented by the shade. 
One can see that the lower bound is $J=\frac{1}{5}$, and the upper bound is $J=1$,
which correspond to $J_{1c}$ and $J_{2c}$, respectively. 

{\it Nearly isotropic nature of half-moon.-}
The energy dispersion of the lower-dispersive band around $\Gamma$ point is isotropic. 
Indeed, $\varepsilon^{\mathrm{K}}_-(\bm{q})$ can be expanded around $\Gamma$ point as 
\begin{align}
\varepsilon^{\mathrm{K}}_-(\bm{q}) \sim & -2(1-J) + \left( \frac{1}{4} - \frac{5J}{4}\right) q^2 + 
\left(- \frac{1}{192} +\frac{17J}{192} \right) q^4 \notag \\
+ & \mathcal{O}(q_i^5), \label{eq:en_expand}
\end{align}
with $q  =\sqrt{q_x^2 +q_y^2}$. 
Equation (\ref{eq:en_expand}) shows that the energy dispersion is isotropic up to the order of $q^4$,
and this leads to nearly circular shape of the energy-minima surface in region (II) [see Fig. \ref{fig:minima_kagome}(a)]. 

\subsection{Pyrochlore lattice \label{sec:appendix_pyrochlore}}
The same method can be applied to a pyrochlore lattice, so here we outline the calculations. 
We consider the exchange matrix on a original pyrochlore lattice:
\begin{equation}
\hat{H}^{\mathrm{P}}(\bm{q}) = \hat{h}^{\mathrm{P}}_1 (\bm{q}) + J \hat{h}^{\mathrm{P}}_2 (\bm{q}), \label{eq:exP}
\end{equation}
where 
\begin{eqnarray}
[\hat{h}^{\mathrm{P}}_1 (\bm{q})]_{\mu \nu} = \left\{
\begin{array}{ll}
2 \cos \bm{q}\cdot (\bm{r}^{\mathrm{P} }_\mu -\bm{r}^{\mathrm{P} }_\nu) & (\mu \neq \nu) \\
0& (\mu = \nu) \\
\end{array}
\right.
\end{eqnarray}
and 
\begin{eqnarray}
[\hat{h}^{\mathrm{P}}_2 (\bm{q})]_{\mu \nu} = \left\{
\begin{array}{ll}
2 \cos \bm{q}\cdot \left( \sum_{\rho \neq \mu, \nu} 
\bm{r}^{\mathrm{P} }_\mu +\bm{r}^{\mathrm{P} }_\nu - 2 \bm{r}^{\mathrm{P} }_\rho \right) & (\mu \neq \nu) \\
2 \cos \bm{q} \cdot \left[2\sum_{\rho \neq \mu} 
 (\bm{r}^{\mathrm{P} }_\mu-\bm{r}^{\mathrm{P} }_\rho) \right]& (\mu = \nu) \\
\end{array}
\right. \nonumber \\
\end{eqnarray}
The polynomial form of the Hamiltonian (\ref{eq:exP}) with respect to $\hat{h}^{\mathrm{P}}_{\mathrm{1} } (\bm{q})$ is obtained as
\begin{equation}
\hat{h}^{\mathrm{P}}_{\mathrm{2} } (\bm{q} ) = [\hat{h}^{\mathrm{P}}_{\mathrm{1} } (\bm{q} )]^2-2\hat{h}^{\mathrm{P}}_{\mathrm{1} } (\bm{q} ) -6\hat{I}_{4\times 4},  \label{eq:pnnn}
\end{equation}
since $z=6$ and $x=2$ for a pyrochlore lattice. 

 Next, the exchange matrix for the dual diamond lattice is given as 
\begin{equation}
\hat{h}^{\mathrm{D}}_{\mathrm{1} } (\bm{q} )
= \left(
\begin{array}{cc}
0& g(\bm{q})\\
g^{\ast}(\bm{q}) &0\\ 
\end{array}
\right), 
\end{equation}
with 
\begin{equation}
g(\bm{q}) =  e^{-i\frac{q_x+q_y+q_z}{4}} + e^{i\frac{q_x+q_y-q_z}{4}} + e^{i\frac{q_x-q_y+q_z}{4}} + e^{i\frac{-q_x+q_y+q_z}{4}}.
\end{equation}
Its eigenvalues are $\varepsilon_{\pm}^{\rm (P)} (\bm{q})  = \pm |g(\bm{q})|$,
and the corresponding eigenvectors are
\begin{equation}
\bm{\psi}^{\mathrm{D}}_\pm(\bm{q}) =  
\frac{1}{\sqrt{2}} 
\left(
\begin{array}{c}
\pm e^{i\theta_g (\bm{q})/2} \\ 
e^{-i\theta_g (\bm{q})/2}
\end{array}
\right),
\end{equation}
with $\theta_g(\bm{q}) = \mathrm{arg}\ g(\bm{q})$. 

Then, using the argument in Sections \ref{sec:linegraph} and \ref{sec:mom},
we obtain the eigenvalues of the Hamiltonian (\ref{eq:exP}) as 
\begin{equation}
\varepsilon^{\mathrm{P}}_\pm (\bm{q}) = J |g(\bm{q})|^2 \pm (1+2J) |g(\bm{q})| + 2 -6J.
\end{equation} 
The corresponding eigenvectors are given as 
\begin{equation}
 \bm{\psi}^{\mathrm{P}}_{\pm}(\bm{q}) = \frac{  h^{\mathrm{P} \leftarrow \mathrm{D} }(\bm{q}) \bm{\psi}^{\mathrm{D}}_\pm(\bm{q})}{|| h^{\mathrm{P} \leftarrow \mathrm{D} }(\bm{q}) \bm{\psi}^{\mathrm{D}}_\pm(\bm{q}) ||}, 
\end{equation}
where $\hat{h}^{\mathrm{P} \leftarrow \mathrm{D}} (\bm{q}) $ is a rectangular matrix 
\begin{equation}
\hat{h}^{\mathrm{P} \leftarrow \mathrm{D}}(\bm{q})
= 
\left(
\begin{array}{cc}
e^{i\phi_1} & e^{-i\phi_1} \\ 
e^{i\phi_2} & e^{-i\phi_2} \\ 
e^{i\phi_3} & e^{-i\phi_3} \\ 
e^{i\phi_4} & e^{-i\phi_4} \\ 
\end{array}
\right),
\end{equation}
and $\phi_1 = \frac{q_x + q_y + q_z}{8}$, $\phi_2=\frac{q_x - q_y - q_z}{8} $, $\phi_3= \frac{-q_x + q_y - q_z}{8}$, and $\phi_4= \frac{-q_x - q_y + q_z}{8}$.
The rest of eigenvectors, i.e. two flat modes, are orthogonal to $\bm{\psi}^{\mathrm{P}, \pm} (\bm{q})$
and their eigenenergy is $-2(1-J)$.

{\it Energy minima.-} 
The minima of $\varepsilon^{\mathrm{P}}_- (\bm{q})$ is obtained by solving
\begin{equation}
\frac{\partial \varepsilon^{\mathrm{P}}_- (\bm{q}) }{\partial q_i }
= [2J |g(\bm{q})| -(1+2J)] \frac{\partial |g( \bm{q})|}{\partial q_i} 
= 0.  \label{eq:min_P}
\end{equation}
(\ref{eq:min_P}) is satisfied when
\begin{enumerate}
\item[(i)] $|g(\bm{q})| = \frac{1+2J}{2J}$,
\item[(ii)] $\frac{\partial |g(\bm{q})|}{\partial q_x}  = \frac{\partial |g(\bm{q})|}{\partial q_y} =\frac{\partial |g(\bm{q})|}{\partial q_z} =0 $.
\end{enumerate}
Again (ii) is satisfied at several high-symmetry points,
which turn out not to be energy minima, 
so let us focus on (i). 
When $0\leq J \leq \frac{1}{6}$, (i) does not have solutions, the lowest-energy band in this region is the flat band. 
Then, for $ \frac{1}{6}  \leq J \leq \frac{1}{2} $, 
the solution is a surface enclosing $\Gamma$ point
[Fig. \ref{fig:minima_pyrochlore}(a)]. 
Finally, for $J \geq \frac{1}{2}$, a solution is a surface enclosing the zone corners 
[Fig. \ref{fig:minima_pyrochlore}(b)]. 
 
{\it Derivation of the phase boundaries.-}
\begin{figure}[t]
\begin{center}
\hspace{-1cm}\includegraphics[bb= 0 0 501 245,width=\linewidth]{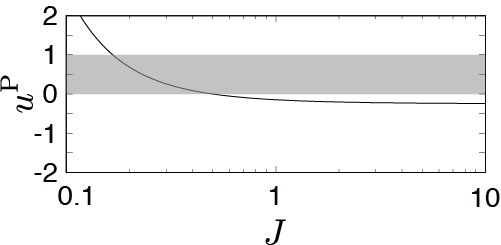}
\vspace{-10pt}
\caption{ $J$ dependence of $u^{\mathrm{P}} \equiv  \frac{-12J^2 +4J-1}{48 J^2}$.
The solution of Eq. (\ref{eq:sol_min_P}) for $0 \leq Q \leq \pi$ exists for 
the shaded area. }
\label{fig:boundary_P}
\end{center}
\end{figure}
Similar to the case of a kagome lattice, 
the phase boundaries for a pyrochlore lattice are determined by the presence/absence of 
the energy minima on $\Gamma$L line. 
On $\Gamma$L line,
the momentum is parametrized as $\bm{q} = (Q,Q,Q) $ with $0\leq Q \leq \pi$. 
Then the condition for the energy minima is given by 
\begin{equation}
|g(\bm{q})| = \sqrt{6\cos Q +10},  \label{eq:sol_min_P}
\end{equation}
and its solution of (i) is
\begin{equation}
Q = 2 \cos^{-1} \left( \frac{-12J^2 +4J-1}{48 J^2} \right).
\end{equation}
We plot $u^{\mathrm{P}} \equiv  \frac{-12J^2 +4J-1}{48 J^2 }$ in Fig. \ref{fig:boundary_P}. 
We again examine the condition that $Q$ is between 0 and $\pi$ (a shaded area of Fig. \ref{fig:boundary_P}),
and find that the lower (upper) bound is $J=\frac{1}{6} \left(\frac{1}{2}\right)$.

\section{Monte Carlo simulations}
\label{app:MC}

Monte Carlo simulations are performed on systems of classical O(3) spins on the kagome and pyrochlore lattices, whose system sizes are respectively $12L^{2}$ and $16L^{3}$ sites. To decorrelate the system, we use jointly the heatbath method, over-relaxation and parallel tempering. Thermalization is made in two steps: first a slow annealing from high temperature to the temperature of measurement $T$ during $t_{e}$ Monte Carlo steps (MCs) followed by $t_{e}$ MCS at temperature $T$. After thermalization, measurements are done every 10 MCs during $t_{m}=10 t_{e}$ MCs. All temperatures are given in units of $J_{1}=1$. The details of each simulation are
as follows:
\begin{itemize}
\item Fig.~\ref{fig:SSF}: $L=30$ for both lattices and $t_{m}=10^{5},10^{6}$ MCs for the pyrochlore and kagome lattice respectively.
\item Fig.~\ref{fig:ChK}, top: $t_{m}=10^{6}$ MCs, and the error bars are coming from an average over $n$ runs with different initial configurations, where $n=50$ for $L<15$ and $n=20$ for $L>15$.
\item Fig.~\ref{fig:ChK}, bottom: $L=20$ and $t_{m}=10^{6}$ MCs.
\item Fig.~\ref{fig:ChSqP}: $L=8$ (a) and $L=16$ (b) and $t_{m}=10^{6}$ MCs. The error bars in (a) are coming from an average over 6 runs with different initial configurations; when not visible, they are smaller than the dots.
\item Fig.~\ref{fig:HMK}: $L=30$ and $t_{m}=10^{5}$ MCs.
\item Fig.~\ref{fig:HMP}: $L=20$ and $t_{m}=10^{5}$ MCs.
\item Fig.~\ref{fig:radius}: $L=30$ and $t_{m}=10^{5}$ MCs.
\item Fig.~\ref{fig:J05obs}: $L\in\{6,8,10\}$ and $t_{m}\in\{10^{7}, 2.10^{7},10^{7}\}$ MCs respectively.
\end{itemize}

\begin{figure}[ht]
\begin{center}
\includegraphics[width=\columnwidth]{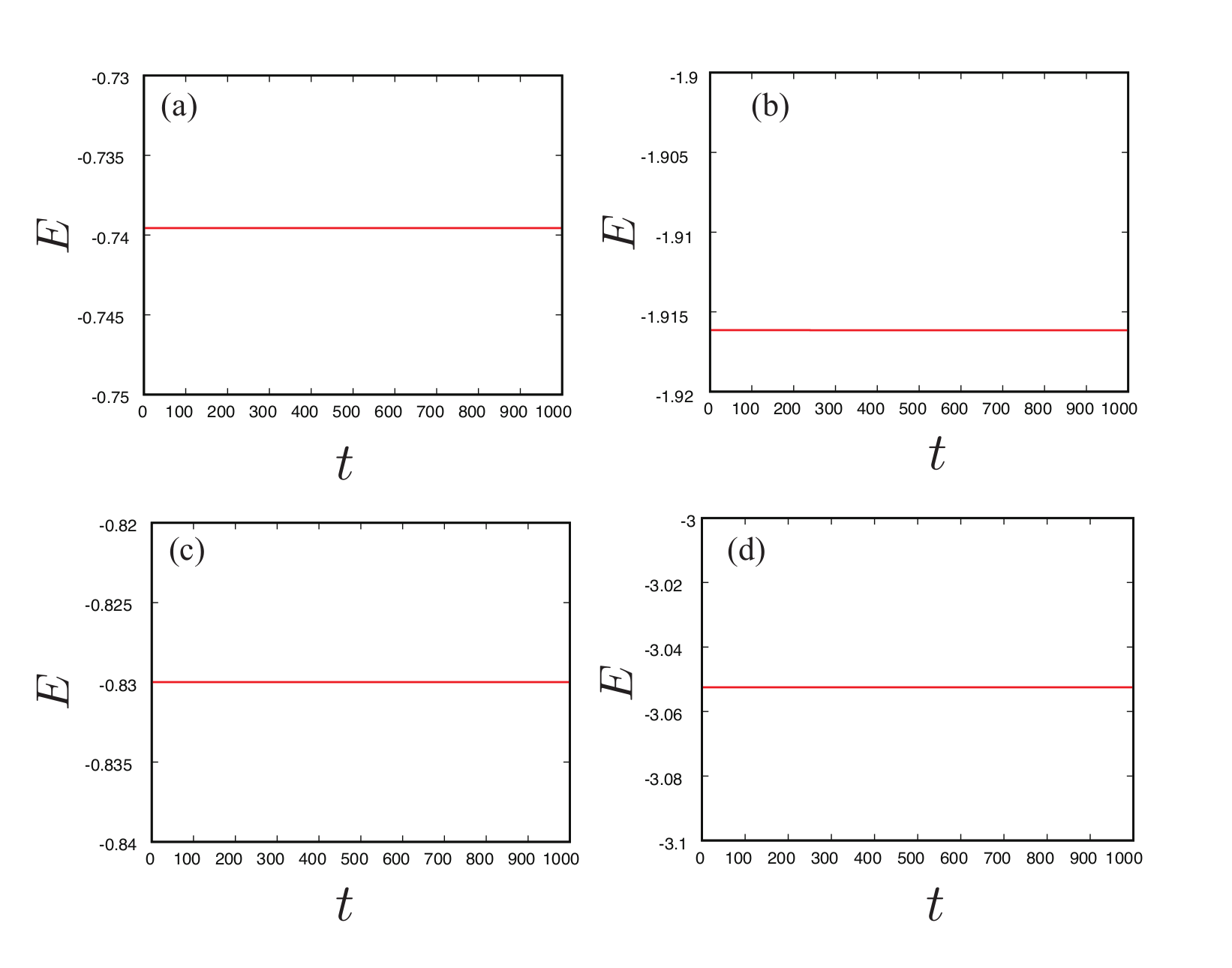}
\caption{
Time evolutions of the energy (per spin) for the fourth-order Runge-Kutta method in the Landau-Lifshitz dynamics for kagome (a), (b) and pyrochlore (c), (d) systems. The conservation of energy is satisfied, within the accuracy of $10^{-5}$.
}
\label{figr1}
\end{center}
\end{figure}

For LL dynamics on the kagome [Fig.~\ref{fig:llg_kagome}] and pyrochlore [Fig.~\ref{fig:llg_pyrochlore}] lattices, we prepared respectively 864  and 432 spin configurations, for system sizes of $3\times 30^2 $ and $4 \times 12^3$ spins. In the parameter region of interest, single-spin-flip Monte Carlo updates were adequate. These spin configurations were then used as seeds for the fourth-order Runge-Kutta method, using $\delta t=0.01$ as the time interval, and $N_t=100000$ as the number of steps of the time evolution.
The accuracy of the numerical simulation was confirmed by calculating the energy; indeed, the energy conservation is satisfied during the Landau-Lifshitz dynamics within the accuracy of $10^{-5}$ (Fig. \ref{figr1}) .

\section{Rank-two tensor order parameter}
\label{app:Q}

The rank-two tensor order parameter is time-reversal invariant and measures the on-site quadrupolar order. For a pyrochlore lattice of $N$ sites, the rank-two tensor is defined following Ref.~\onlinecite{shannon10}:
\begin{eqnarray}
\mathcal{Q}^{\alpha}=\dfrac{1}{N}\sum_{i=1}^{N}\mathcal{Q}_{i}^{\alpha},
\end{eqnarray}
where
\begin{eqnarray}
\mathcal{Q}^{3z^{2}-r^{2}}_{i}&=&\dfrac{1}{\sqrt{3}}\left[2 (S_{i}^{z})^{2} - (S_{i}^{x})^{2} - (S_{i}^{y})^{2}\right],\\
\mathcal{Q}^{x^{2}-y^{2}}_{i}&=&(S_{i}^{x})^{2} - (S_{i}^{y})^{2},\\
\mathcal{Q}^{xy}_{i}&=&2 S_{i}^{x}\;S_{i}^{y},\\
\mathcal{Q}^{yz}_{i}&=&2 S_{i}^{y}\;S_{i}^{z},\\
\mathcal{Q}^{zx}_{i}&=&2 S_{i}^{z}\;S_{i}^{x}.
\end{eqnarray}
The order parameter used in Fig.~\ref{fig:J05obs}.(b) comes from the norm of all quadrupole moments
\begin{eqnarray}
M_Q=\sqrt{\sum_{\alpha}\left(\mathcal{Q}^{\alpha}\right)^{2}}\;,
\label{eq:Q}
\end{eqnarray}
and is saturated when all spins are collinear, taking the value $2/\sqrt{3}$.

\section{Origin of subextensive entropy in the ordered state at $J=J_{2c}$ for pyrochlore \label{app:order} }
In this appendix, we explain the origin of the sub-extensive entropy of (1/2,1/2,1/2)-state at $J=J_{2c}$, illustrated in Fig.~\ref{fig:meth}. For this state, which consists of the double-charge clusters, the energy of the ground state comes from the contact between charges [Eq.~(\ref{eq:EGSJ05})]. These contacts only take place within each cluster. Let us recall there is one double-charge cluster per cubic unit cell, with one double charge, four single charges, and three vacuum tetrahedra. It means that any change which does not affect the integrity and connectivity of the clusters is iso-energetic. Thanks to the vacuum surrounding every cluster, such changes are possible by shifting an entire plane of clusters. An example is given in Fig.~\ref{fig:meth}. The plane of clusters in the upper part of the figure is shifted in the [110] direction between the left and right panels, while the bottom part remains fixed -- the thickness of a plane is exactly one cubic unit cell. More precisely all spins along the [110] lines are shifted by a distance of two nearest neighbors; the spins along the [$1\bar{1}0$] lines are left unchanged. Thanks to the intervening layers of vacuum tetrahedra (the one just below the transparent plane, and the one at the top of the figure), this shift does not affect any NN pair of charges. The resulting state is thus also a ground state. A second shift in the [110] direction gives back the initial state. The same is also true if one shifts the spins in the upper plane along the [$1\bar{1}0$] lines, leaving the [110] lines unchanged. The addition of both shifts makes a fourth possibility. This gives $4^{L}$ ground states for a system of $L$ planes in the [001] direction.

The same reasoning applies for planes orthogonal to the [010] and [100] directions. However, it is not possible to do successively a shift in a (100) plane followed by a shift in a (001) plane. This is because the conservation of the energy depends on the intervening vacuum layers. Visually, a plane of clusters can glide at no energy cost as long as there is a layer of vacuum to isolate it from the two planes above and below. But, the shift of a (100) plane intersects the layers of vacuum orthogonal to the [001] direction; the shift of a (001) plane is now forbidden in the ground state. As a result, the entropy of the ground state is sub-extensive, of the order of $L\ln 4$.

The configurations of Fig.~\ref{fig:meth}, as well as the shift of entire planes, have been observed in snapshots of Monte Carlo simulations at low temperatures, up to fluctuations away from collinearity. The presence of the shifting planes favors a given cubic axis but not a given direction. 
This spontaneously breaks rotational symmetry as measured by the quadrupolar order parameter $M_{Q}$. But, in this picture, the correlations along the preferred cubic axis should be ``paramagnetic'' and the order parameter $M_{L}$ should vanish like $1/L$ in the thermodynamic limit. This is not what is observed in simulations. The reason is because the system is not made of Ising spins, but of continuous Heisenberg spins. The ground state is selected via order by disorder because of soft modes around the $L$ points of the Brillouin zone, which favor the long-range ordered states where none of the planes are shifted. However, the sub-extensive entropy is probably responsible for the difficulty of the simulations to thermalize at very low temperatures.



\begin{thebibliography}{1}
\bibitem{cap1}
Here the upward/downward is defined with respect to 
$y$-direction for a kagome lattice, and $(111)$-directions of a pyrochlore lattice.
\end{thebibliography}
\end{document}